\documentclass[aps,preprint,showpacs,preprintnumbers,amsmath,amssymb]{revtex4}
\usepackage{amsmath,mathrsfs,amsbsy,color,graphicx,bm,amsthm,amsfonts}
\usepackage{units}
\usepackage{bbm}
\usepackage{braket}%
\usepackage{times}
\usepackage{dcolumn}
\usepackage{mathrsfs}
\usepackage{amsmath,amssymb,epsfig}
\usepackage[svgnames]{xcolor}
\usepackage{amsmath}
\usepackage{amssymb}
\usepackage{tikz}
\usetikzlibrary{arrows.meta, calc, shapes.geometric}

\definecolor{journalblue}{RGB}{41, 128, 185}   
\definecolor{journalorange}{RGB}{230, 126, 34} 
\definecolor{journalgreen}{RGB}{46, 204, 113}  
\definecolor{journalgray}{RGB}{149, 165, 166}   
\definecolor{warmgold}{HTML}{E6A848} 
\definecolor{inkblack}{HTML}{121212}
\definecolor{rindlerpurple}{RGB}{147, 98, 190}  
\definecolor{rindlerred}{RGB}{235, 93, 95}       
\definecolor{rindleryellow}{RGB}{232, 175, 71}   
\definecolor{rindlerbrown}{RGB}{177, 134, 91}    
\newcommand{\myhbar}[1]{%
  \mkern 2mu%
  \vbox{%
    \hrule height 0.8pt%
    \kern 1.8pt%
    \hbox{$#1$}%
  }%
  \mkern 2mu%
}
\newcommand{\Dbar}{\myhbar{D}}
\begin{document}

\title{Causal-diamond thermalization induces nonseparability in N-partite quantum systems}
\author{ Hui-Chen Yang, Shu-Min Wu\footnote{Email: smwu@lnnu.edu.cn (corresponding authors)} }
\affiliation{Department of Physics, Liaoning Normal University, Dalian 116029, China}


\begin{abstract}
We investigate the nonseparability of multipartite bosonic and fermionic $GHZ$ and $W$ states in a causal diamond spacetime using the Abe-Rajagopal (AR) $q$-conditional entropy. The finite lifetime of the observer gives rise to a causal diamond horizon, which induces an Unruh-like thermal effect and leads to a nontrivial restructuring of nonseparability in $N$-partite systems. Our main result is that the thermal effect  can enhance a net nonseparability of fermionic $W$ states, in sharp contrast to the general expectation that relativistic thermalization leads to a monotonic degradation of bosonic nonseparability. In addition, we find that fermionic  nonseparability is generally more robust than its bosonic counterpart under causal diamond restrictions. Among different entangled resources, $GHZ$ states exhibit stronger nonseparability and greater robustness than $W$ states under identical causal conditions. We further show that the nonseparability of $W$ states decreases with increasing particle number $N$, whereas that of $GHZ$ states remains independent of $N$  in causal diamond spacetime. These results demonstrate that particle statistics, entanglement structure, and observer lifetime jointly determine the persistence of nonseparability in causally restricted spacetimes, providing insights for relativistic quantum information processing.
\end{abstract}

\vspace*{0.5cm}
\maketitle
\section{Introduction}
In relativistic quantum field theory, the causal structure of spacetime fundamentally influences how vacuum fluctuations and quantum correlations are perceived by different observers \cite{SDF1,SDF2,SDF3,SDF4,SDF5,SDF6,SDF7,SDF8,SDF9,SDF10,SDF11,SDF12}. Besides black-hole and Rindler horizons, causal diamonds provide a particularly simple framework for investigating horizon-induced quantum phenomena. A causal diamond corresponds to the spacetime region accessible to an observer with a finite lifetime and is bounded by an effective causal horizon \cite{SDF60}, despite the absence of spacetime curvature or acceleration. Quantization with respect to diamond modes reveals that the Minkowski vacuum is a two-mode squeezed state between the interior and exterior regions. Tracing over the inaccessible exterior modes therefore produces a thermal state with a temperature inversely proportional to the observer's lifetime \cite{SDF61}. This thermal response originates entirely from causal restriction rather than spacetime dynamics, making causal diamonds an ideal setting for isolating the effects of horizons on quantum systems. Consequently, causal diamonds have emerged as a promising platform in relativistic quantum information, particularly for exploring the behavior of multipartite quantum correlations under causally restricted structures. Their potential applications extend to the study of quantum correlations, vacuum entanglement harvesting, and the interplay between generalized entropy and gravitational physics in many-body quantum systems.

In recent years, quantum information in relativistic settings, particularly under causal restrictions, has developed into an active interdisciplinary field at the interface of quantum information theory, quantum field theory, and spacetime physics \cite{SDF13,SDF14,SDF15,SDF16,SDF17,SDF18,SDF19,SDF20,SDF21,SDF22,SDF23,SDF24,SDF25,SDF26,SDF27,SDF28,SDF29,SDF30,SDF31,SDF32,SDF33,SDF34,SDF35,SDF36,SDF37,SDF38,SDF39,SDF40,SDF41,SDF42,SDF43,SDF44,SDF45,SDF46,SDF47,SDF48,SDF49,SDF50,SDF51,SDF52,SDF53,SDF54,
SDF55,SDF56,SDF57,SDF58,SDF59,QFL1,QFL2,QFL3}. Extensive studies have investigated the impact of horizon-induced effects on bipartite quantum resources, including entanglement, quantum steering, quantum discord, coherence, quantum Fisher information, and entropic uncertainty relations, in various relativistic backgrounds such as black hole spacetimes, Rindler wedges  \cite{SDF1,SDF2,SDF3,SDF4,SDF5,SDF6,SDF7,SDF8,SDF9,SDF10,SDF11,SDF12}, and, more recently, causal diamond spacetime  \cite{SDF60,SDF61}. In these scenarios, the restriction of accessible field modes by causal boundaries typically induces effective thermalization and leads to degradation of quantum correlations. With the rapid progress of quantum technologies toward large-scale implementations. However, bipartite and tripartite entangled states are no longer sufficient to capture the complexity required for advanced quantum information tasks. Recent experimental advances have already demonstrated the controllable preparation of large-scale quantum systems. For example, Bluvstein \emph{et al.} demonstrated a fault-tolerant neutral-atom quantum computing architecture based on reconfigurable arrays of up to 448 neutral atoms \cite{SDF62}, while Gao \emph{et al.} demonstrated the superconducting quantum processor Zuchongzhi~3, containing 105 readable qubits and 182 couplers \cite{SDF63}.  It is therefore of both fundamental and practical importance to extend relativistic quantum information studies to multipartite nonseparable states under causal constraints. In causal diamond spacetime, the finite lifetime of the observer induces a nontrivial decomposition of field modes into causally accessible and inaccessible sectors, and tracing over the inaccessible region transforms the global pure state into a mixed density matrix, thereby potentially altering the separability structure of multipartite correlations. However, a systematic analysis of multipartite entanglement in this setting is technically challenging due to the highly nontrivial mode mixing induced by the diamond decomposition and the exponential growth of Hilbert space dimension with particle number, which makes analytic evaluation of entanglement measures difficult, especially for structured states such as generalized $W$ states. To overcome these difficulties, we employ the AR $q$-conditional entropy as a diagnostic tool to characterize  nonseparability of  N-partite quantum systems in causal diamond backgrounds. Compared with standard criteria based on von Neumann conditional entropy, the AR formalism provides stricter separability bounds and serves as a more sensitive probe of multipartite quantum correlations under causally induced thermalization.

Building on these considerations, we investigate the effects of causal-diamond-induced thermalization on the nonseparability of $N$-partite $GHZ$ and $W$ states for free bosonic and fermionic fields using the AR $q$-conditional entropy. We consider a scenario in which $N$ observers initially share multipartite entangled states in the Minkowski vacuum, while one observer is confined to a finite-lifetime causal diamond and the remaining observers remain inertial. The causal restriction partitions the field into accessible and inaccessible diamond modes, and tracing over the exterior region gives rise to an effective thermal state that modifies the nonseparability of the shared quantum system. We derive analytical expressions for the nonseparability of both bosonic and fermionic fields and identify several distinctive features  in a causal
diamond spacetim. Most notably, causal-diamond thermalization can enhance the nonseparability of the fermionic $W$ state, in sharp contrast to the monotonic degradation observed for bosonic fields. Moreover, fermionic states are generally more robust than their bosonic counterparts, while $GHZ$ states retain stronger nonseparability than $W$ states under identical causal restrictions. We further show that the nonseparability of the $W$ state decreases with increasing particle number under causal-diamond thermalization, whereas that of the $GHZ$ state remains independent of the system size. These results demonstrate that particle statistics, entanglement structure, and causal restrictions jointly determine the persistence of quantum nonseparability  in finite-lifetime spacetimes, providing new insights into relativistic quantum information under causal horizons.

The remainder of this paper is organized as follows. In Sec. II, we review the AR $q$-conditional entropy used to characterize the nonseparability of multipartite quantum systems. Section III introduces the quantization of free bosonic and fermionic fields in causal diamond spacetime. In Sec. IV, we investigate the nonseparability of $N$-partite $GHZ$ and $W$ states under causal-diamond-induced thermalization. Finally, we conclude in Sec. V.

\section{Quantification of nonseparability in a quantum system}
In this section, we briefly review the AR $q$-conditional entropy, which extends the Tsallis non-extensive statistical framework to the characterization of correlations in a quantum systems \cite{SDF64,SDF65,SDF66,SDF67}. Consider a composite system consisting of two subsystems $X$ and $Y$, described by a joint probability distribution $p_{ij}(X,Y)$, which specifies the probability that subsystem $X$ occupies its $i$-th state while subsystem $Y$ is simultaneously in its $j$-th state. The corresponding marginal and conditional probabilities follow from standard Bayesian relations, thereby encoding the statistical structure of the composite system.

Within the Tsallis non-extensive formalism, the uncertainty of a quantum state $\rho$ is quantified by the Tsallis $q$-entropy, which generalizes the Shannon entropy and is defined as
\begin{equation}
S_q^{(T)}(\rho)=\frac{\mathrm{Tr}\left(\rho^{\,q}\right)-1}{1-q}.
\label{eq:Tsallis}
\end{equation}
This entropy is nonnegative and satisfies a generalized composition rule of the form
\begin{equation}
S_q^{(T)}(\rho_{XY})
=
S_q^{(T)}(\rho_X)
+
S_q^{(T)}(\rho_Y)
+
(1-q)\,S_q^{(T)}(\rho_X)\,S_q^{(T)}(\rho_Y),
\end{equation}
which explicitly reflects its intrinsic nonadditive structure. However, this factorization property is generally violated in the presence of nonlocal quantum correlations, such as entanglement, where the composite system cannot be described in terms of independent subsystems.
To capture such correlations, Abe and Rajagopal introduced the $q$-conditional entropy, defined as
\begin{equation}
S_q(X|Y)
=
\frac{
S_q^{(T)}(\rho_{XY})
-
S_q^{(T)}(\rho_Y)
}{
1+(1-q)\,S_q^{(T)}(\rho_Y)
}.
\label{eq:AR_def}
\end{equation}
Substituting Eq.~\eqref{eq:Tsallis} into Eq.~\eqref{eq:AR_def}, one obtains an equivalent spectral representation,
\begin{equation}
S_q(X|Y)
=
\frac{1}{q-1}
\left[
1-
\frac{
\mathrm{Tr}\left( \rho_{XY}^{\,q} \right)
}{
\mathrm{Tr}\left( \rho_Y^{\,q} \right)
}
\right]
=
\frac{1}{q-1}
\left[
1-
\frac{
\sum_n \lambda_n^{\,q}(\rho_{XY})
}{
\sum_m \lambda_m^{\,q}(\rho_Y)
}
\right],
\label{eq:AR_spectral}
\end{equation}
where $\{\lambda_n\}$ and $\{\lambda_m\}$ denote the eigenvalues of the density matrix $\rho_{XY}$ and its reduced state $\rho_Y$, respectively. This form preserves the generalized nonadditivity of Tsallis entropy and provides a direct spectral characterization of conditional uncertainty in the subsystem $X$.
A central property of the AR $q$-conditional entropy is that it remains nonnegative for all separable bipartite states. In contrast, for entangled states, quantum correlations can suppress $\mathrm{Tr}(\rho_{XY}^{\,q})$ relative to $\mathrm{Tr}(\rho_Y^{\,q})$, leading to the possibility of negative values of $S_q(X|Y)$. Therefore, the condition $S_q(X|Y) < 0$ provides a sufficient criterion for nonseparability, which reduces to the standard von Neumann conditional entropy criterion in the limit $q \to 1$. Moreover, increasing the parameter $q$ typically strengthens the separability constraint, with the limit $q \to \infty$ providing the most restrictive spectral bound.

\section{Quantization of bosonic and fermionic fields in diamond spacetime}
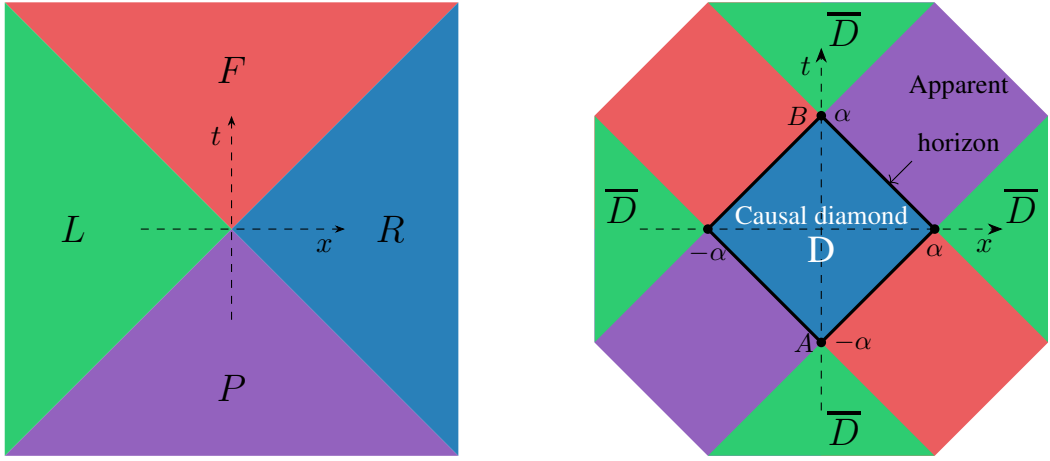
\begin{figure}[htbp]
\centering
\begin{tikzpicture}[scale=3.0, font=\small]

    \begin{scope}[shift={(0,0)}]
        \fill[rindlerred] (0,0) -- (1,1) -- (-1,1) -- cycle;    
        \fill[rindlerpurple]  (0,0) -- (1,-1) -- (-1,-1) -- cycle;  
        \fill[journalgreen]    (0,0) -- (-1,1) -- (-1,-1) -- cycle;  
        \fill[journalblue] (0,0) -- (1,1) -- (1,-1) -- cycle;    
        
        \draw[-{Stealth[scale=0.8]}, dashed, thin] (0,-0.4) -- (0,0.5) node[left, pos=0.9] {$t$};
        \draw[-{Stealth[scale=0.8]}, dashed, thin] (-0.4,0) -- (0.5,0) node[below, pos=0.9] {$x$};
        
        \node at (0, 0.7) {\large $F$};
        \node at (0, -0.7) {\large $P$};
        \node at (-0.7, 0) {\large $L$};
        \node at (0.7, 0) {\large $R$};
    \end{scope}

    \begin{scope}[shift={(2.6,0)}] 
        
        \def\maxL{1.0}     
        \def\alphaVal{0.5} 

        \begin{scope}
            \clip (-\maxL+\alphaVal, \maxL) -- (\maxL-\alphaVal, \maxL) -- 
                  (\maxL, \maxL-\alphaVal) -- (\maxL, -\maxL+\alphaVal) -- 
                  (\maxL-\alphaVal, -\maxL) -- (-\maxL+\alphaVal, -\maxL) -- 
                  (-\maxL, -\maxL+\alphaVal) -- (-\maxL, \maxL-\alphaVal) -- cycle;
            
            \fill[rindlerred] (-\maxL, 0) rectangle (0, \maxL);
            \fill[rindlerpurple] (0, 0) rectangle (\maxL, \maxL);
            \fill[rindlerpurple] (-\maxL, -\maxL) rectangle (0, 0);
            \fill[rindlerred] (0, -\maxL) rectangle (\maxL, 0);
            
            \fill[journalgreen] (-\maxL, \maxL-\alphaVal) -- (-\alphaVal, 0) -- (-\maxL, -\maxL+\alphaVal) -- cycle;
            \fill[journalgreen] (\maxL, \maxL-\alphaVal) -- (\alphaVal, 0) -- (\maxL, -\maxL+\alphaVal) -- cycle;
            
            \fill[journalgreen] (-\maxL+\alphaVal, \maxL) -- (0, \alphaVal) -- (\maxL-\alphaVal, \maxL) -- cycle;
            \fill[journalgreen] (-\maxL+\alphaVal, -\maxL) -- (0, -\alphaVal) -- (\maxL-\alphaVal, -\maxL) -- cycle;
        \end{scope}

        \fill[journalblue] (0, \alphaVal) -- (\alphaVal, 0) -- (0, -\alphaVal) -- (-\alphaVal, 0) -- cycle;
        
        \draw[very thick] (0, \alphaVal) -- (\alphaVal, 0) -- (0, -\alphaVal) -- (-\alphaVal, 0) -- cycle;

        \draw[thin, dashed, -{Stealth[length=2mm]}] (-0.8, 0) -- (0.8, 0) node[below, pos=0.95] {$x$};
        \draw[thin, dashed, -{Stealth[length=2mm]}] (0, -0.8) -- (0, 0.8) node[left, pos=0.95] {$t$};

        \node at (0, -0.09) {\large \textcolor{white}{D}};
        \node[font=\footnotesize] at (0, 0.06) {\textcolor{white}{Causal diamond}};
        
        \node[font=\large] at (0.1, 0.88) {$\Dbar$};
        \node[font=\large] at (0.1, -0.88) {$\Dbar$};
        \node[font=\large] at (-0.88, 0.1) {$\Dbar$};
        \node[font=\large] at (0.88, 0.1) {$\Dbar$};

        \fill (0, \alphaVal) circle (0.6pt);
        \fill (0, -\alphaVal) circle (0.6pt);
        \fill (\alphaVal, 0) circle (0.6pt);
        \fill (-\alphaVal, 0) circle (0.6pt);
        
        \node[left, font=\footnotesize, xshift=-1pt] at (0, \alphaVal) {$B$};
        \node[right, font=\footnotesize, xshift=1pt] at (0, \alphaVal) {$\alpha$};
        
        \node[left, font=\footnotesize, xshift=1pt] at (0, -\alphaVal) {$A$};
        \node[right, font=\footnotesize, xshift=1pt] at (0, -\alphaVal) {$-\alpha$};
        
        \node[below, font=\footnotesize, yshift=-2pt] at (\alphaVal, 0) {$\alpha$};
        \node[below, font=\footnotesize, yshift=-2pt] at (-\alphaVal, 0) {$-\alpha$};

        \draw[<-] (\alphaVal*0.6, \alphaVal*0.4) -- (0.4, 0.3) node[above right, font=\footnotesize, align=center, xshift=-5pt] {Apparent\\horizon};
    \end{scope}

\end{tikzpicture}
\caption{Diagrammatic correspondence between Rindler wedges and causal diamond under the composite conformal map. Left panel: Standard Rindler wedges ($R,L,F,P$) in Minkowski-like coordinates. Right panel: Transformed diamond geometry, with interior domain $D=\{(x,t):|t|+|x|\le\alpha\}$ (blue square) and four disjoint exterior domains $\bar D$. The boundaries of $D$ are the apparent horizons, with past/future tips at birth $A(t=-\alpha)$ and death $B(t=\alpha)$ events. Colors indicate the one-to-one matching of regions across the transformation.}
\label{fig:spacetime_mapping}
\end{figure}

The causal diamond for a finite-lifetime observer is defined as the intersection of the future light cone of the birth event and the past light cone of the death event in  Fig.~\ref{fig:spacetime_mapping} \cite{SDF68,SDF69,SDF70,SDF71}. The observer's proper lifetime is 
$T=2\alpha$, and the causally accessible region is the diamond $D := \{(x,t): |t| + |x| \le \alpha\}.$ Coordinates adapted to the causal diamond are constructed via its conformal relation to Rindler spacetime. The finite diamond region $D$ can be mapped one-to-one onto the right Rindler wedge $R=\{(\tilde t,\tilde x):|\tilde t|\le \tilde x,\;\tilde x\ge0\}$, allowing the Rindler coordinate system to be transferred to the diamond \cite{SDF72,SDF73}. Since conformal transformations preserve causal structure, this mapping provides a natural description of both the interior and exterior regions. A two-step composite map unifies these transformations into a single systematic framework \cite{SDF61}.

The first step selects an appropriate combination  of special conformal transformations \(K(\rho)\), dilation scalings \(\Lambda(\lambda)\), and translations \(T(\alpha)\) to generate a composite, conformal, one-to-one mapping $(\tilde t, \tilde x)\to (t,x)$ from the right Rindler wedge $\mathrm{R}$ to the diamond region $\mathrm{D}$, described by the Minkowski coordinates \((t,x)\) \cite{SDF74}. Introducing the parameter \(\lambda\), one defines the rescaled parameter $\tilde{\alpha} = \frac{2\alpha}{\lambda}.$
The composite transformation
$[T(-\alpha)\circ K(1/2\alpha)\circ \Lambda(\lambda)]$ leads to the explicit coordinate relations
\begin{equation}
\frac{t}{\alpha} =
\frac{2\,(\tilde t/\tilde\alpha)}
{(\tilde x/\tilde\alpha + 1)^2 - (\tilde t/\tilde\alpha)^2},
\qquad
\frac{x}{\alpha} =
\frac{-1 - (\tilde x/\tilde\alpha)^2 + (\tilde t/\tilde\alpha)^2}
{(\tilde x/\tilde\alpha + 1)^2 - (\tilde t/\tilde\alpha)^2},
\end{equation}
with inverse transformations
\begin{equation}\label{S0}
\frac{\tilde t}{\tilde\alpha} =
\frac{2\,(t/\alpha)}
{(x/\alpha - 1)^2 - (t/\alpha)^2},
\qquad
\frac{\tilde x}{\tilde\alpha} =
\frac{1 - (x/\alpha)^2 + (t/\alpha)^2}
{(x/\alpha - 1)^2 - (t/\alpha)^2}.
\end{equation}
Scale consistency ensures that all physical observables, including the diamond temperature, are independent of $\lambda$. In the second step, the right wedge is parametrized by Rindler coordinates $(\eta,\xi)$  and mapped to $(\tilde t,\tilde x)$ via the rescaled Rindler transformation \cite{SDF75,SDF76,SDF77}. With parameter $\lambda$ and sign $\varepsilon=\pm1$ distinguishing interior and exterior regions, the mapping reads
\begin{equation}
\frac{\tilde t}{\tilde\alpha} =
\varepsilon\, e^{2\xi/\alpha} \sinh\!\left(\frac{2\eta}{\alpha}\right),
\qquad
\frac{\tilde x}{\tilde\alpha} =
\varepsilon\, e^{2\xi/\alpha} \cosh\!\left(\frac{2\eta}{\alpha}\right),
\end{equation}
where \(\eta, \xi \in (-\infty,\infty)\).  Curves with constant \(\xi\) correspond to observers undergoing uniform acceleration
$
a(\xi) = \frac{2}{\alpha e^{2\xi/\alpha}},
$
with the reference value \(a = 2/\alpha\) at \(\xi = 0\). For exterior regions, mapped from the Rindler future and past wedges, time and space coordinates are locally interchanged ($\tilde t \leftrightarrow \tilde x$), with the full expressions obtained by sign changes and permutations. Combining both steps establishes a one-to-one correspondence among Rindler, intermediate, and diamond coordinates, allowing both interior and exterior of the diamond to be represented by $(\eta,\xi)$. Although the intermediate conformal map contains the parameter $\lambda$, scale consistency ensures that all physical quantities are independent of $\lambda$. To simplify algebra and highlight propagation directions ($\sigma=\pm1$ for left/right movers), we switch to null coordinates.
Define the Minkowski null variables
\begin{equation}\label{S1}
U_\sigma = t+\sigma x,\qquad \tilde U_\sigma = \tilde t+\sigma \tilde x,\qquad u_\sigma = \varepsilon(\eta+\sigma\xi),
\end{equation}
with $\varepsilon=\pm1$ ensuring future-directed null coordinates. For $\sigma=\pm1$, these yield the standard light-cone coordinates: in Minkowski, $U_+ \equiv V=t+x$, $U_- \equiv U=t-x$; in Rindler, $\tilde U_+ \equiv \tilde V=t_R+x_R$, $\tilde U_- \equiv \tilde U=t_R-x_R$; and in the diamond, $u_+ \equiv v=\varepsilon(\eta+\xi)$, $u_- \equiv u=\varepsilon(\eta-\xi)$. With these definitions, the transformation between the intermediate ($\tilde U_\sigma$) and diamond ($U_\sigma$) null coordinates takes the compact form
\begin{equation}\label{S2}
\frac{\tilde V}{\tilde\alpha} = \frac{1+V/\alpha}{1-V/\alpha},\qquad 
\frac{\tilde U}{\tilde\alpha} = -\frac{1-U/\alpha}{1+U/\alpha},
\end{equation}
which is equivalent to Eq.~(\ref{S0}).

In these coordinates, the combined conformal and Rindler transformations take particularly simple forms. For the interior diamond region $D$, one finds
\begin{equation}\label{s2}
e^{2v/\alpha}
=
\frac{1+V/\alpha}{1-V/\alpha},
\qquad
e^{2u/\alpha}
=
\frac{1+U/\alpha}{1-U/\alpha}.
\end{equation}
For the exterior region $\bar D$, the corresponding relations become
\begin{equation}\label{s3}
e^{2\bar v/\alpha}
=
\frac{V/\alpha-1}{V/\alpha+1},
\qquad
e^{2\bar u/\alpha}
=
\frac{U/\alpha-1}{U/\alpha+1}.
\end{equation}
Eqs. (\ref{s2}) and (\ref{s3}) can be inverted straightforwardly. For the interior diamond $D$, one obtains
\begin{equation}\label{s4}
\frac{V}{\alpha}
=
\tanh\!\left(\frac{v}{\alpha}\right),
\qquad
\frac{U}{\alpha}
=
\tanh\!\left(\frac{u}{\alpha}\right),
\end{equation}
while for the exterior region $\bar D$,
\begin{equation}\label{s5}
\frac{V}{\alpha}
=
-\coth\!\left(\frac{\bar v}{\alpha}\right),
\qquad
\frac{U}{\alpha}
=
-\coth\!\left(\frac{\bar u}{\alpha}\right).
\end{equation}
These expressions explicitly relate the Minkowski null coordinates to the corresponding diamond null coordinates in the interior and exterior regions.

\subsection{Bosonic Field}


The quantization of a massless scalar field in the causal diamond is formulated based on its geometric structure. Since the massless Klein-Gordon equation is conformally invariant in $(1+1)$-dimensional spacetime, the conformal transformation between Minkowski and diamond coordinates leaves the field equation unchanged. Consequently, a complete orthonormal set of positive-frequency mode functions can be constructed in both the interior and exterior regions of the causal diamond, providing the basis for the canonical quantization of the scalar field.

In Minkowski coordinates $(t,x)$, the massless scalar field satisfies the Klein-Gordon equation
\begin{equation}
(\partial_t^2-\partial_x^2)\Phi
=\partial_U\partial_V\Phi=0,
\end{equation}
where $U=t-x$ and $V=t+x$. A complete set of positive frequency Minkowski modes is given by
\begin{equation}
f_{+,k}(V)=\frac{1}{4\pi k}e^{-ikV},\qquad
f_{-,k}(U)=\frac{1}{4\pi k}e^{-ikU}.
\end{equation}
To describe the field from the perspective of a causal diamond, we introduce the diamond coordinates $(\eta,\xi)$, with $v=\varepsilon(\eta+\xi)$ and $u=\varepsilon(\eta-\xi)$, which are related to the Minkowski null coordinates through Eqs. (\ref{S1}). Similarly, in diamond coordinates, the Klein-Gordon equation reads
\begin{equation}
(\partial_\eta^2-\partial_\xi^2)\Phi
=
\partial_v\partial_u\Phi
=
0.
\end{equation}
The corresponding normalizable mode solutions can be separated into interior and exterior diamond modes
\begin{equation}
g_{\sigma,\omega}^{B,(\mathrm{int})}(U_\sigma)
=\frac{1}{\sqrt{{4\pi\omega}}}\,
e^{-i\omega u_\sigma(U_\sigma)}\,\theta(\alpha-|U_\sigma|),
\end{equation}
\begin{equation}																			g_{\sigma,\omega}^{B,(\mathrm{ext})}(U_\sigma)
=\frac{1}{\sqrt{{4\pi\omega}}}\,
e^{-i\omega u_\sigma(U_\sigma)}\,\theta(|U_\sigma|-\alpha),
\end{equation}
where $B$ denotes the bosonic field. The index $\sigma=\pm1$ labels the left- and right-moving sectors, $\omega>0$ denotes the mode frequency, and $\theta(x)$ is the Heaviside step function. The null coordinate $u_{\sigma}(U_{\sigma})$ is defined by the coordinate transformations in Eqs.~\eqref{S1}-\eqref{s5}, while the Heaviside step functions ensure that the mode functions have support only in the interior $(|U_{\sigma}|<\alpha)$ and exterior $(|U_{\sigma}|>\alpha)$ regions of the causal diamond, respectively. The interior and exterior mode functions together constitute a complete orthonormal basis for the expansion of the scalar field operator in causal diamond spacetime. Consequently, the field operator can be expanded as
\begin{equation}
\Phi
=
\sum_{\sigma=\pm}
\int_0^\infty d\omega
\left[
b_{\sigma,\omega}^{B,(\mathrm{int})}
g_{\sigma,\omega}^{B,(\mathrm{int})}
+
b_{\sigma,\omega}^{B,(\mathrm{ext})}
g_{\sigma,\omega}^{B,(\mathrm{ext})}
+\mathrm{H.c.}
\right],
\label{eq17}
\end{equation}
where H.c. denotes the Hermitian conjugate. The operators
$b_{\sigma,\omega}^{B,(\mathrm{int})}$ and
$b_{\sigma,\omega}^{B,(\mathrm{ext})}$ satisfy the canonical bosonic commutation relations
\begin{equation}
\left[
b_{\sigma,\omega}^{B,(\mathrm{int})},
b_{\sigma',\omega'}^{B,(\mathrm{int})\dagger}
\right]
=
\delta_{\sigma\sigma'}
\delta(\omega-\omega'),
\qquad
\left[
b_{\sigma,\omega}^{B,(\mathrm{ext})},
b_{\sigma',\omega'}^{B,(\mathrm{ext})\dagger}
\right]
=
\delta_{\sigma\sigma'}
\delta(\omega-\omega'),
\label{eq18}
\end{equation}
while all other commutators vanish.

Following the analytic continuation procedure for the construction of Unruh-diamond modes \cite{SDF61}, the interior and exterior diamond modes can be combined into Unruh-diamond modes containing only positive Minkowski-frequency components. The corresponding Bogoliubov coefficients are parametrized by the squeezing parameter $r_{\omega}$, satisfying $\tanh r_\omega = e^{-\pi\alpha\omega/2}.$ This construction leads to the explicit expressions for the Unruh-diamond modes
\begin{equation}\label{hB1}
h_{\sigma,\omega}^{B,(\mathrm{int})}
=\frac{1}{\sqrt{1-e^{-\pi\alpha\omega}}}\,
g_{\sigma,\omega}^{B,(\mathrm{int})}
+\frac{e^{-\pi\alpha\omega/2}}{\sqrt{1-e^{-\pi\alpha\omega}}}\,
g_{\sigma,\omega}^{B,(\mathrm{ext})*},
\end{equation}
\begin{equation}\label{hB2}
h_{\sigma,\omega}^{B,(\mathrm{ext})}
=\frac{1}{\sqrt{1-e^{-\pi\alpha\omega}}}\,
g_{\sigma,\omega}^{B,(\mathrm{ext})}
+\frac{e^{-\pi\alpha\omega/2}}{\sqrt{1-e^{-\pi\alpha\omega}}}\,
g_{\sigma,\omega}^{B,(\mathrm{int})*}.
\end{equation}
From the mode relations in Eqs.~\eqref{hB1} and \eqref{hB2}, one obtains the corresponding Bogoliubov transformations between the annihilation operators
\begin{equation}
c_{\sigma,\omega}^{B,(\mathrm{int})}
=\frac{1}{\sqrt{1-e^{-\pi\alpha\omega}}}\,
b_{\sigma,\omega}^{B,(\mathrm{int})}
-\frac{e^{-\pi\alpha\omega/2}}{\sqrt{1-e^{-\pi\alpha\omega}}}\,
b_{\sigma,\omega}^{B,(\mathrm{ext})\dagger},
\end{equation}
\begin{equation}
c_{\sigma,\omega}^{B,(\mathrm{ext})}
=\frac{1}{\sqrt{1-e^{-\pi\alpha\omega}}}\,
b_{\sigma,\omega}^{B,(\mathrm{ext})}
-\frac{e^{-\pi\alpha\omega/2}}{\sqrt{1-e^{-\pi\alpha\omega}}}\,
b_{\sigma,\omega}^{B,(\mathrm{int})\dagger},
\end{equation}
where $b^{B,(\mathrm{int})}_{\sigma,\omega}$ and $b^{B,(\mathrm{ext})}_{\sigma,\omega}$ are the annihilation operators associated with the interior and exterior diamond modes, respectively. These Bogoliubov transformations characterize how a
finite-lifetime observer confined to the causal diamond perceives the
Minkowski vacuum. Since the Unruh-diamond vacuum is annihilated by both
the internal and external mode operators, imposing the corresponding
vacuum conditions yields the following explicit two-mode squeezed state
\begin{equation}\label{BSLT}
|0_{\sigma,\omega}\rangle_{U}^{B}
=\sqrt{1-e^{-\pi\alpha\omega}}
\sum_{n=0}^\infty e^{-\frac{1}{2}n\pi\alpha\omega}
|n\rangle_{\mathrm{int}}\otimes|n\rangle_{\mathrm{ext}},
\end{equation}
and single-particle excited state
\begin{equation}\label{BSYT}
|1_{\sigma,\omega}\rangle_{U}^{B}
=({1-e^{-\pi\alpha\omega}})\,
\sum_{n=0}^\infty
\sqrt{n+1}\,e^{-\frac{1}{2}n\pi\alpha\omega}\,
|n+1\rangle_{\mathrm{int}}\otimes|n\rangle_{\mathrm{ext}}
\end{equation}
where
$\{|n\rangle_{(\mathrm{int})}\}$ and
$\{|n\rangle_{(\mathrm{ext})}\}$ represent the orthonormal Fock bases
associated with the interior and exterior regions of the causal diamond,
respectively. Since a finite-lifetime observer is confined to the
interior of the causal diamond, the exterior modes are inaccessible and
must be traced out. Consequently, the observer perceives the Minkowski
vacuum as a thermal state, with the corresponding particle number
distribution given by
\begin{equation}
 N_{\omega}^{B}
=
\frac{1}{e^{\pi\alpha\omega}-1},
\label{BoseDistribution}
\end{equation}
which is the Bose-Einstein distribution at the diamond temperature $T_D=\frac{1}{\pi\alpha}$.

\subsection{Fermionic Field}
Similar to the bosonic field, the Bogoliubov transformations relating the
Unruh diamond operators to the interior and exterior diamond operators for the
fermionic field can be expressed as
\begin{equation}
c_{\sigma,\omega}^{F,\text{(int)}} = \frac{1}{\sqrt{e^{-\pi\alpha\omega} + 1}} b_{\sigma,\omega}^{F,\text{(int)}} - \frac{1}{\sqrt{e^{\pi\alpha\omega} + 1}} b_{\sigma,\omega}^{F,\text{(ext)}\dagger} , \label{eq:28}
\end{equation}
\begin{equation}
c_{\sigma,\omega}^{F,\text{(ext)}} = \frac{1}{\sqrt{e^{-\pi\alpha\omega} + 1}} b_{\sigma,\omega}^{F,\text{(ext)}} + \frac{1}{\sqrt{e^{\pi\alpha\omega} + 1}} b_{\sigma,\omega}^{F,\text{(int)}\dagger} , \label{eq:29}
\end{equation}
where $F$ denotes the fermionic field  \cite{SDF60}. Consequently, the Unruh diamond vacuum state and the excited states of the
fermionic field in diamond spacetime can be written as
\begin{equation}\label{f-jt}
|0\rangle_{U}^{F} =
\frac{1}{\sqrt{e^{-\pi\alpha\omega}+1}}\,
|0\rangle_{\text{int}}|0\rangle_{\text{ext}}
+
\frac{1}{\sqrt{e^{\pi\alpha\omega}+1}}\,
|1\rangle_{\text{int}}|1\rangle_{\text{ext}},
\end{equation}
\begin{equation}\label{f-jft}
|1\rangle_{U}^{F}
=
|1\rangle_{\text{int}}\,|0\rangle_{\text{ext}}.
\end{equation}
The corresponding expectation value of the particle number for the mode of frequency $\omega$ is
\begin{equation}\label{FMN}
N^{F}_{\omega}
=
\frac{1}{e^{\pi\alpha\omega}+1},
\end{equation}
which corresponds to a thermal Fermi-Dirac distribution. From Eqs.~\eqref{BoseDistribution} and \eqref{FMN}, it is evident that the Bose-Einstein and
Fermi-Dirac statistics governing bosonic and fermionic fields,
respectively, lead to fundamentally different thermal responses in
causal diamond spacetime. These statistical distinctions manifest themselves through different evolutions of nonseparability under the diamond thermal effect, highlighting the crucial role of particle statistics in determining the resilience and dynamical behavior of nonseparable states perceived by finite-lifetime observers.

\section{Nonseparability of Bosonic and Fermionic Fields in Causal Diamond Spacetime}
In this section, we investigate the nonseparability of multipartite bosonic and fermionic states in causal diamond spacetime using the AR $q$-conditional entropy. We consider $N$ observers sharing multipartite $GHZ$ and $W$ states and characterize their nonseparability under the bipartition between the causally unrestricted and causally restricted subsystems.
Initially, we assume that $N$ ($N\ge3$) observers prepare either a $GHZ$ state or a $W$ state in the global Minkowski vacuum, corresponding to the idealized situation in which all observers have infinite lifetimes and unrestricted access to the field modes,
\begin{equation}
\lvert \mathrm{GHZ}^{B/F}_{123\cdots N} \rangle
= \frac{1}{\sqrt{2}}
\left(
\lvert 0 0 \cdots 0 \rangle
+
\lvert 1 1 \cdots 1 \rangle
\right),
\end{equation}
\begin{equation}\label{WBF}
\lvert W^{B/F}_{123\cdots N} \rangle
= \frac{1}{\sqrt{N}}
\left(
\lvert 1 0 \cdots 0 \rangle
+
\lvert 0 1 \cdots 0 \rangle
+
\cdots
+
\lvert 0 0 \cdots 1 \rangle
\right),
\end{equation}
where $i$ ($i=1,2,\ldots,N$) labels the field mode detected by observer $\mathcal{O}_i$, while the superscripts $B$ and $F$ denote bosonic and fermionic fields, respectively.
We then assume that one observer has a finite proper lifetime $T=2\alpha$ and is therefore confined to a causal diamond, whereas the remaining $N-1$ observers remain in the infinite-lifetime limit. The presence of the diamond horizon requires the corresponding field mode to be decomposed into interior and exterior diamond modes. Since the exterior modes are causally inaccessible to the finite-lifetime observer, they are traced out, yielding a reduced density matrix defined on the diamond interior. The resulting mixed state provides the basis for evaluating the nonseparability of multipartite bosonic and fermionic systems under causal-diamond-induced thermalization.

\subsection{Nonseparability of the bosonic field in causal diamond spacetime}

By employing the Unruh-diamond transformations that relate Minkowski modes to diamond interior and exterior modes given in Eqs.~\eqref{BSLT} and ~\eqref{BSYT} , the $N$-partite bosonic $GHZ$ state can be rewritten in causal diamond spacetime as
\begin{align}
\lvert \mathrm{GHZ}^{B}_{123\cdots N+1} \rangle
&= \frac{1}{\sqrt{2}}
\Bigg[
\left(1-e^{-\pi \alpha \omega}\right)^{\frac{1}{2}}\overbrace{ \left(|0\rangle_1|0\rangle_2\cdots|0\rangle_{N-1}\right)}^{\ket{\bar{0}}}
\sum_{n=0}^{\infty}
e^{-\frac{1}{2}n\pi\alpha\omega}
\lvert n \rangle_{N,\mathrm{int}}
\lvert n \rangle_{N+1,\mathrm{ext}}
\nonumber\\%
&+
\left(1-e^{-\pi \alpha \omega}\right)
\overbrace{ \left(|1\rangle_1|1\rangle_2\cdots|1\rangle_{N-1}\right)}^{\ket{\bar{1}}}
\sum_{m=0}^{\infty}
e^{-\frac{1}{2}m\pi\alpha\omega}
\sqrt{m+1}\,
\lvert m+1 \rangle_{N,\mathrm{int}}
\lvert m \rangle_{N+1,\mathrm{ext}}
\Bigg],
\end{align}
where
$
\lvert \bar{0} \rangle
= \lvert 0 \rangle_{1}\lvert 0 \rangle_{2}\cdots \lvert 0 \rangle_{N-1}$ and $
\lvert \bar{1} \rangle
= \lvert 1 \rangle_{1}\lvert 1 \rangle_{2}\cdots \lvert 1 \rangle_{N-1}.
$
In causal diamond spacetime, the interior and exterior regions are causally disconnected due to the presence of the diamond horizons. Consequently, the modes supported in the exterior region are inaccessible to an observer confined within the diamond. Tracing over the exterior modes, we obtain the reduced mixed density matrix for the diamond interior region as
\begin{align}
\rho^{B,\mathrm{GHZ}}_{123\cdots N_\mathrm{int}}
&= \frac{1}{2}
\sum_{n=0}^{\infty}
e^{-n\pi\alpha\omega}
\bigg[
\bigl(1-e^{-\pi\alpha\omega}\bigr)
\lvert \bar{0} \rangle\langle \bar{0} \rvert
\otimes
\lvert n \rangle_{N,\mathrm{int}}\langle n \rvert
\nonumber\\
&+
\bigl(1-e^{-\pi\alpha\omega}\bigr)^{\frac{3}{2}}
\sqrt{n+1}\,
\lvert \bar{0} \rangle\langle \bar{1} \rvert
\otimes
\lvert n \rangle_{N,\mathrm{int}}\langle n+1 \rvert
\nonumber\\
&+
\bigl(1-e^{-\pi\alpha\omega}\bigr)^{\frac{3}{2}}
\sqrt{n+1}\,
\lvert \bar{1} \rangle\langle \bar{0} \rvert
\otimes
\lvert n+1 \rangle_{N,\mathrm{int}}\langle n \rvert
\nonumber\\
&+
\bigl(1-e^{-\pi\alpha\omega}\bigr)^{2}
(n+1)\,
\lvert \bar{1} \rangle\langle \bar{1} \rvert
\otimes
\lvert n+1 \rangle_{N,\mathrm{int}}\langle n+1 \rvert
\bigg].
\end{align}
The resulting density matrix exhibits a block-diagonal structure composed of independent \(2\times2\) blocks along the diagonal, while all other off-diagonal elements vanish. It can be represented compactly as
\begin{align}
\rho^{B,\mathrm{GHZ}}_{123\cdots N_\mathrm{int}}
=
\frac{1}{2}
\begin{pmatrix}
0 & & & & \\
&\Lambda_{0} & & & \\
& & \Lambda_{1} & & \\
& & & \ddots & \\
& & & & \Lambda_{n} & \\
& & & &  & \ddots
\end{pmatrix},
\end{align}
where each block \(\Lambda_{n}\) takes the form
\begin{align}
\Lambda_{n}\!\left(\rho^{B,\mathrm{GHZ}}_{123\cdots N_\mathrm{int}}\right)
=
\begin{pmatrix}
(1-\gamma)\gamma^{n}
&
(1-\gamma)^{\frac{3}{2}}\sqrt{n+1}\,\gamma^{n}
\\[6pt]
(1-\gamma)^{\frac{3}{2}}\sqrt{n+1}\,\gamma^{n}
&
(1-\gamma)^{2}(n+1)\gamma^{n}
\end{pmatrix},
\end{align}
with $\gamma = e^{-\pi\alpha\omega}$.
The eigenvalues of the $n$-th block of the reduced density matrix
$\rho^{B,\mathrm{GHZ}}_{123\cdots N_\mathrm{int}}$, obtained after tracing over the diamond exterior modes, are given by $0$ and
\begin{equation}\label{41}
\lambda^{B}_{\mathrm{GHZ}_n}
=
\frac{1}{2}
\left(1-e^{-\pi\alpha\omega}\right)
e^{-n\pi\alpha\omega}
\left[2+n-(n+1)e^{-\pi\alpha\omega}\right].
\end{equation}
It is straightforward to verify that the density matrix is properly normalized, since $\mathrm{Tr}\!\left[\rho^{B,\mathrm{GHZ}}_{123\cdots N_\mathrm{int}}\right] = \sum_{n=0}^{\infty}\lambda^{B}_{\mathrm{GHZ},n} = 1$.

To further quantify the nonseparability, we construct the reduced density matrix $\rho^{B,\mathrm{GHZ}}_{N_\mathrm{int}}$, corresponding to the modes of the last qubit inside the causal diamond.
This density matrix is obtained by tracing out the first $N-1$ subsystems,
$\rho^{B,\mathrm{GHZ}}_{N_\mathrm{int}} = \mathrm{Tr}_{123\cdots\,N-1}
\!\left[ \rho^{B,\mathrm{GHZ}}_{123\cdots N_\mathrm{int}} \right]$,
which yields
\begin{equation}
\rho^{B,\mathrm{GHZ}}_{N_\mathrm{int}}
=
\frac{1}{2}
\left(1-e^{-\pi\alpha\omega}\right)
\sum_{m=0}^{\infty}
e^{-m\pi\alpha\omega}
\left[
1+m\left(1-e^{-\pi\alpha\omega}\right)e^{\pi\alpha\omega}
\right]
|m\rangle_{N,\mathrm{int}}\langle m| .
\end{equation}
This reduced density matrix is diagonal in the interior Fock basis and infinite dimensional. Its $m$-th eigenvalue is given by
\begin{equation}
\lambda^{B}_{\mathrm{GHZ}_m}
=
\frac{1}{2}
\left(1-e^{-\pi\alpha\omega}\right)
e^{-m\pi\alpha\omega}
\left[
1+m\left(1-e^{-\pi\alpha\omega}\right)e^{\pi\alpha\omega}
\right].
\end{equation}
Again, the normalization condition $\mathrm{Tr}\!\left[\rho^{B,\mathrm{GHZ}}_{N_\mathrm{int}}\right] = \sum_{m=0}^{\infty}\lambda^{B}_{\mathrm{GHZ}_m} = 1$ is satisfied.

To characterize the nonseparability of the $N$-partite bosonic $GHZ$ state in causal diamond spacetime, we substitute the eigenvalues $\lambda^{B}_{\mathrm{GHZ}_n}$ of $\rho^{B,\mathrm{GHZ}}_{123\cdots N_\mathrm{int}}$ and $\lambda^{B}_{\mathrm{GHZ}_m}$ of $\rho^{B,\mathrm{GHZ}}_{N_\mathrm{int}}$ into the AR $q$-conditional entropy defined in Eq.~\eqref{eq:AR_spectral}, obtaining
\begin{equation}\label{BGHZ1}
S^{\mathrm{GHZ},N}_{B,q}
=
\frac{1}{q-1}
\left\{
1-
\frac{
\displaystyle
\sum_{n=0}^{\infty}
\left\{
e^{-n\pi\alpha\omega}
\left[
2+n-(n+1)e^{-\pi\alpha\omega}
\right]
\right\}^q
}{
\displaystyle
\sum_{m=0}^{\infty}
\left\{
e^{-m\pi\alpha\omega}
\left[
1+m\left(1-e^{-\pi\alpha\omega}\right)e^{\pi\alpha\omega}
\right]
\right\}^q
}
\right\}.
\end{equation}
Negative values of the AR $q$-conditional entropy $S^{B,q}_{\mathrm{GHZ},N}$ signal the nonseparability between the single causally restricted interior subsystem and the remaining $N-1$ causally unrestricted subsystems of the bosonic $GHZ$ state.
From Eq.~\eqref{BGHZ1}, it is evident that the Unruh-diamond effect, encoded in the finite-lifetime parameter $\alpha$,
modifies the statistical weights of the interior modes and hence influences the degree of bosonic nonseparability.
Importantly, the  nonseparability of the bosonic $GHZ$ state is completely independent of the total number of subsystems $N$. This invariance indicates that the diamond-induced thermal spectrum does not alter the qualitative nonseparability behavior when additional qubits remain in the causally unrestricted region, highlighting a key structural distinction from other multipartite configurations such as the $W$ state.

Similarly, the pure $N$-partite bosonic $W$ state can be expressed in terms of Unruh-diamond modes as
\begin{align}
\lvert W^{B}_{123\cdots N+1}\rangle
&=\frac{1}{\sqrt{N}}\Bigg[
(1-e^{-\pi\alpha\omega})^{\frac{1}{2}}\,
\lvert 1\rangle_1\lvert 0\rangle_2\cdots\lvert 0\rangle_{N-1}
\sum_{n=0}^{\infty}e^{-\frac{1}{2}n\pi\alpha\omega}
\lvert n\rangle_{N,\mathrm{int}}\lvert n\rangle_{N+1,\mathrm{ext}}
\nonumber\\
&+
(1-e^{-\pi\alpha\omega})^{\frac{1}{2}}\,
\lvert 0\rangle_1\lvert 1\rangle_2\cdots\lvert 0\rangle_{N-1}
\sum_{n=0}^{\infty}e^{-\frac{1}{2}n\pi\alpha\omega}
\lvert n\rangle_{N,\mathrm{int}}\lvert n\rangle_{N+1,\mathrm{ext}}
\nonumber\\
&+\cdots+
(1-e^{-\pi\alpha\omega})^{\frac{1}{2}}\,
\lvert 0\rangle_1\lvert 0\rangle_2\cdots\lvert 1\rangle_{N-1}
\sum_{n=0}^{\infty}e^{-\frac{1}{2}n\pi\alpha\omega}
\lvert n\rangle_{N,\mathrm{int}}\lvert n\rangle_{N+1,\mathrm{ext}}
\nonumber\\
&+
(1-e^{-\pi\alpha\omega})
\lvert \bar{0}\rangle
\sum_{n=0}^{\infty}\sqrt{n+1}e^{-\frac{1}{2}n\pi\alpha\omega}
\lvert n+1\rangle_{N,\mathrm{int}}\lvert n\rangle_{N+1,\mathrm{ext}}
\Bigg].
\end{align}
After tracing over the exterior degrees of freedom, we obtain the reduced density operator $\rho^{B,W}_{123\cdots N_\mathrm{int}}$ as
\begin{equation}
\rho^{B,W}_{123\cdots N_\mathrm{int}}
=\frac{1}{N}\bigl(\rho^{B}_{\mathrm{diag}}+\rho^{B}_{\mathrm{off}}\bigr),
\end{equation}
where the diagonal contribution $\rho^{B}_{\mathrm{diag}}$ is given by
\begin{align}
\rho^{B}_{\mathrm{diag}}&=
\sum_{i=1}^{N-1}
\lvert 1\rangle_i\langle 1\rvert
\bigotimes_{\substack{j=1\\ j\neq i}}^{N-1}
\lvert 0\rangle_j\langle 0\rvert\,
(1-e^{-\pi\alpha\omega})
\sum_{n=0}^{\infty}e^{-n\pi\alpha\omega}
\lvert n\rangle_{N,\mathrm{int}}\langle n\rvert
\nonumber\\
&+
\bigotimes_{i=1}^{N-1}\lvert 0\rangle_i\langle 0\rvert\,
(1-e^{-\pi\alpha\omega})^{2}
\sum_{n=0}^{\infty}(n+1)e^{-n\pi\alpha\omega}
\lvert n+1\rangle_{N,\mathrm{int}}\langle n+1\rvert ,
\end{align}
and the off-diagonal part $\rho^{B}_{\mathrm{off}}$ reads
\begin{align}
\rho^{B}_{\mathrm{off}}&=
\sum_{\substack{i,j=1\\ i\neq j}}^{N-1}
\lvert 0\rangle_i\langle 1\rvert\,
\lvert 1\rangle_j\langle 0\rvert
\bigotimes_{\substack{k=1\\ k\neq i,j}}^{N-1}
\lvert 0\rangle_k\langle 0\rvert\,
(1-e^{-\pi\alpha\omega})
\sum_{n=0}^{\infty}e^{-n\pi\alpha\omega}
\lvert n\rangle_{N,\mathrm{int}}\langle n\rvert
\nonumber\\
&+
\sum_{i=1}^{N-1}
\lvert 1\rangle_i\langle 0\rvert
\bigotimes_{\substack{j=1\\ j\neq i}}^{N-1}
\lvert 0\rangle_j\langle 0\rvert\,
(1-e^{-\pi\alpha\omega})^{\frac{3}{2}}
\sum_{n=0}^{\infty}\sqrt{n+1}\,e^{-n\pi\alpha\omega}
\lvert n\rangle_{N,\mathrm{int}}\langle n+1\rvert
\nonumber\\
&+
\sum_{i=1}^{N-1}
\lvert 0\rangle_i\langle 1\rvert
\bigotimes_{\substack{j=1\\ j\neq i}}^{N-1}
\lvert 0\rangle_j\langle 0\rvert\,
(1-e^{-\pi\alpha\omega})^{\frac{3}{2}}
\sum_{n=0}^{\infty}\sqrt{n+1}\,e^{-n\pi\alpha\omega}
\lvert n+1\rangle_{N,\mathrm{int}}\langle n\rvert .
\end{align}
The resulting density matrix $\rho^{B,W}_{123\cdots N_\mathrm{int}}$ exhibits a block-diagonal structure composed of independent subblocks along the diagonal, while all remaining off-block elements vanish. It can be written schematically as
\begin{equation}
\rho^{B,W}_{123\cdots N_\mathrm{int}}
=\frac{1}{N}
\begin{pmatrix}
0 & & & & \\
&\Lambda'_0 & & & \\
& & \Lambda'_1 & & \\
& & & \ddots & \\
& & & & \Lambda'_n & \\
& & & &  & \ddots
\end{pmatrix},
\end{equation}
where the diagonal blocks $\Lambda'_n$ of the reduced density matrix in causal diamond spacetime are represented by
\begin{equation}
\begin{split}
&\Lambda'_n\!\left(\rho^{B,W}_{123\cdots N_\mathrm{int}}\right)
=\\
&\gamma^{n}
\begin{pmatrix}
(1-\gamma) & (1-\gamma) & \cdots & (1-\gamma) & (1-\gamma)^{\frac{3}{2}}\sqrt{n+1} \\
(1-\gamma) & (1-\gamma) & \cdots & (1-\gamma) & (1-\gamma)^{\frac{3}{2}}\sqrt{n+1} \\
\vdots & \vdots & \ddots & \vdots & \vdots \\
(1-\gamma) & (1-\gamma) & \cdots & (1-\gamma) & (1-\gamma)^{\frac{3}{2}}\sqrt{n+1} \\
(1-\gamma)^{\frac{3}{2}}\sqrt{n+1} &
(1-\gamma)^{\frac{3}{2}}\sqrt{n+1} &
\cdots &
(1-\gamma)^{\frac{3}{2}}\sqrt{n+1} &
(1-\gamma)^2 (n+1)
\end{pmatrix},
\end{split}
\end{equation}
which is a block matrix defined in the subspace
$
\{\,|10\cdots0n\rangle,\; |01\cdots0n\rangle,\;\ldots\; |00\cdots1n\rangle,\; |00\cdots0n+1\rangle \,\}.
$
The matrix $\Lambda'_n\!\left(\rho^{B,W}_{123\cdots N_\mathrm{int}}\right)$ possesses a characteristic structure: it is an $N\times N$ matrix in which the first $N-1$ rows are identical, each containing repeated elements proportional to $(1-\gamma)\gamma^{n}$, except for the last column, which is weighted by $(1-\gamma)^{\frac{3}{2}}\sqrt{n+1}\,\gamma^{n}$. The last row consists of $(1-\gamma)^{\frac{3}{2}}\sqrt{n+1}\,\gamma^{n}$ in the first $N-1$ columns, while the bottom-right element is given by $(1-\gamma)^2 (n+1)\gamma^{n}$.
As a consequence of this highly symmetric block structure, the spectrum of each $n$-th block contains a single nonvanishing eigenvalue, while all remaining eigenvalues are zero. The nonzero eigenvalue of the $n$-th block of the state $\rho^{B,W}_{123\cdots N_\mathrm{int}}$ can be written as
\begin{equation}
\lambda^{B,W}_{n}
=
\frac{1}{N}\,
e^{-n\pi\alpha\omega}
\left(1-e^{-\pi\alpha\omega}\right)
\left[
N+n-(n+1)e^{-\pi\alpha\omega}
\right].
\end{equation}
The reduced density matrix of the $N$-th subsystem inside the causal diamond, $\rho^{B,W}_{N_\mathrm{int}}$, is obtained by tracing out the remaining $N-1$ subsystems. This yields
\begin{equation}
\begin{split}
\rho^{B,W}_{N_\mathrm{int}}
&=
\frac{1}{N}
\sum_{m=0}^{\infty}
e^{-m\pi\alpha\omega}
\left(1-e^{-\pi\alpha\omega}\right)
\Bigl[
\left(1-e^{-\pi\alpha\omega}\right)(m+1)\,
|m+1\rangle_{N,\mathrm{int}}\langle m+1|\\
&+
(N-1)\,|m\rangle_{N,\mathrm{int}}\langle m|
\Bigr],
\end{split}
\end{equation}
from which the $m$-th eigenvalue of $\rho^{B,W}_{N_\mathrm{int}}$ is readily obtained as
\begin{equation}
\lambda^{B,W}_{m}
=
\frac{1}{N}\,
e^{-m\pi\alpha\omega}
\left(1-e^{-\pi\alpha\omega}\right)
\left[
N-1
+
m\,e^{\pi\alpha\omega}\left(1-e^{-\pi\alpha\omega}\right)
\right].
\end{equation}
Substituting the eigenvalues $\lambda^{B,W}_{n}$ of $\rho^{B,W}_{123\cdots N_\mathrm{int}}$ and
$\lambda^{B,W}_{m}$ of $\rho^{B,W}_{N_\mathrm{int}}$ into the AR $q$-conditional entropy, we obtain
\begin{equation}\label{SBW}
S^{W,N}_{B,q}
=
\frac{1}{q-1}
\left\{
1
-
\frac{\displaystyle
\sum_{n=0}^{\infty}
\big\{
e^{-n\pi\alpha\omega}
\left[
N+n-(n+1)e^{-\pi\alpha\omega}
\right]
\big\}^{q}}
{\displaystyle
\sum_{m=0}^{\infty}
\big\{
e^{-m\pi\alpha\omega}
\left[
N-1
+
m\,e^{\pi\alpha\omega}\left(1-e^{-\pi\alpha\omega}\right)
\right]
\big\}^{q}}
\right\}.
\end{equation}
From Eq.~\eqref{SBW}, it follows that  the nonseparability of the bosonic $W$ state depends explicitly on the initial particle number $N$ in causal diamond spacetime. This behavior contrasts with that of the $GHZ$ state, whose nonseparability remains independent of $N$ under causal-diamond
thermalization.

\begin{figure}
\begin{minipage}[t]{0.5\linewidth}
\centering
\includegraphics[width=3.0in,height=5.2cm]{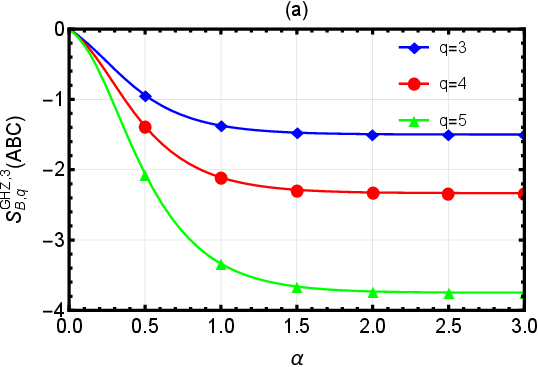}
\end{minipage}%
\begin{minipage}[t]{0.5\linewidth}
\centering
\includegraphics[width=3.0in,height=5.2cm]{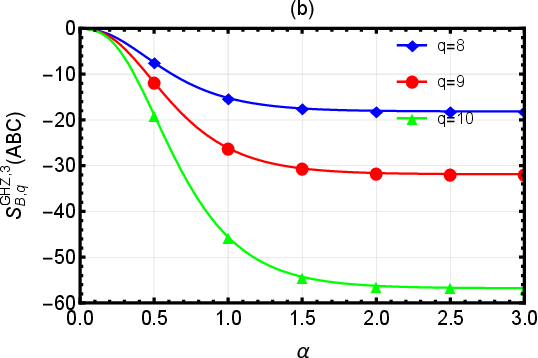}
\end{minipage}%
\vspace{0.5cm}
\begin{minipage}[t]{0.5\linewidth}
\centering
\includegraphics[width=3.0in,height=5.2cm]{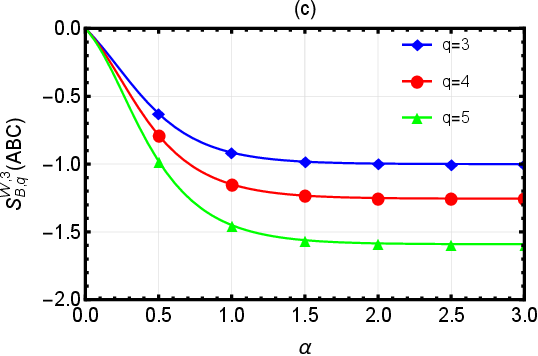}
\end{minipage}%
\begin{minipage}[t]{0.5\linewidth}
\centering
\includegraphics[width=3.0in,height=5.2cm]{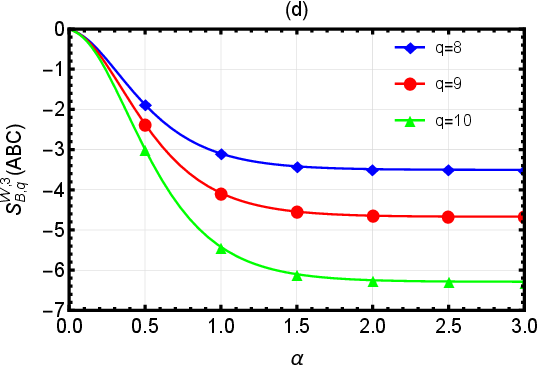}
\end{minipage}%
\caption{AR $q$-conditional entropy for bosonic $GHZ$ and W states in causal diamond spacetime, plotted as a function of the observer’s finite-lifetime parameter $\alpha$ for different values of $q$, with $\omega = 1$ and $N = 3$.}\label{fig:B-GHZ-W-3}
\end{figure}

In Fig.~\ref{fig:B-GHZ-W-3}, we plot the AR $q$-conditional entropy 
$S^{\mathrm{GHZ},3}_{B,q}(ABC)$ and $S^{W,3}_{B,q}(ABC)$ as a function of the finite-lifetime parameter $\alpha$ for different values of $q$, obtained from Eqs.~\eqref{BGHZ1} and \eqref{SBW}. Since more negative values of the AR $q$-conditional entropy correspond to stronger nonseparability, the monotonic decrease of these quantities with increasing $\alpha$ indicates that both bosonic $GHZ$ and $W$ states become increasingly nonseparable as the observer lifetime grows. This behavior can be attributed to the reduction of the causal-diamond temperature, $T_D=1/(\pi\alpha)$, which weakens the thermal fluctuations induced by the diamond horizon and thereby preserves quantum correlations more effectively. For a fixed value of $\alpha$, the AR $q$-conditional entropy of the $W$ state is consistently larger than that of the $GHZ$ state, revealing that the $GHZ$ state possesses stronger nonseparability under identical causal restrictions. This difference originates from the distinct entanglement structures of the two states, which respond differently to the thermalization induced by the causal diamond. These results demonstrate that the observer's lifetime and the initial structure of quantum resources jointly determine the robustness of  nonseparability in causal diamond spacetime.

\subsection{Nonseparability of the fermionic field in causal diamond spacetime}
Similar to the bosonic $GHZ$ state, we use the Unruh-diamond modes defined in Eqs.~\eqref{f-jt} and \eqref{f-jft} to rewrite the fermionic $GHZ$ state as
\begin{equation}
\begin{split}
\vert \text{GHZ}_{123\cdots N+1}^F \rangle &= \frac{1}{\sqrt{2}} \Bigg\{
\overbrace{(\vert 0 \rangle_1 \vert 0 \rangle_2 \cdots \vert 0 \rangle_{N-1})}^{\ket{\bar{0}}} \Big[
(1+e^{-\pi \alpha \omega})^{-\frac{1}{2}} \vert 0 \rangle_{N,\text{int}} \vert 0 \rangle_{N+1,\text{ext}}\\
&+ (1+e^{\pi \alpha \omega})^{-\frac{1}{2}} \vert 1 \rangle_{N,\text{int}} \vert 1 \rangle_{N+1,\text{ext}}
\Big]
+ \overbrace{(\vert 1 \rangle_1 \vert 1 \rangle_2 \cdots \vert 1 \rangle_{N-1})}^{\ket{\bar{1}}} \vert 1 \rangle_{N,\text{int}} \vert 0 \rangle_{N+1,\text{ext}}
\Bigg\}.
\end{split}
\end{equation}
After tracing over the inaccessible exterior diamond modes, we obtain the reduced density operator $\rho_{123\cdots N_\text{int}}^{F,\text{GHZ}}$ as
\begin{equation}
\begin{split}
\rho_{123\cdots N_\text{int}}^{F,\text{GHZ}} &= \frac{1}{2} \Bigg\{
\vert \bar{0} \rangle \langle \bar{0} \vert \Big[ (1+e^{-\pi \alpha \omega})^{-1} \vert 0 \rangle_{N,\text{int}} \langle 0 \vert + (1+e^{\pi \alpha \omega})^{-1} \vert 1 \rangle_{N,\text{int}} \langle 1 \vert \Big]\\
&+   \vert \bar{0} \rangle \langle \bar{1} \vert (1+e^{-\pi \alpha \omega})^{-\frac{1}{2}}\vert 0 \rangle_{N,\text{int}} \langle 1 \vert + \vert \bar{1} \rangle \langle \bar{0} \vert(1+e^{-\pi \alpha \omega})^{-\frac{1}{2}} \vert 1 \rangle_{N,\text{int}} \langle 0 \vert \\
&+ \vert \bar{1} \rangle \langle \bar{1} \vert \vert 1 \rangle_{N,\text{int}} \langle 1 \vert
\Bigg\}.
\end{split}
\end{equation}
The eigenvalues of the reduced density matrix $\rho_{123\cdots N_\text{int}}^{F,\text{GHZ}}$ read
\begin{equation}
\label{eq:eigenvalues_ghz_fermion_1}
\lambda_{\text{GHZ}_1}^F = \frac{1}{2(1+e^{\pi \alpha \omega})}, \quad
\lambda_{\text{GHZ}_2}^F = \frac{1+2e^{\pi \alpha \omega}}{2(1+e^{\pi \alpha \omega})}.
\end{equation}
To further characterize  nonseparability, we construct the reduced density matrix corresponding to the $N$-th interior subsystem by tracing out the remaining $N-1$ subsystems. This yields
\begin{equation}
\label{eq:rho_N_int_ghz_fermion}
\rho_{N_\text{int}}^{F,\text{GHZ}} = \frac{1}{2} \bigg\{
(1+e^{-\pi \alpha \omega}) \vert 0 \rangle_{N,\text{int}} \langle 0 \vert
+ \Big[ 1 + (1+e^{\pi \alpha \omega})^{-1} \Big] \vert 1 \rangle_{N,\text{int}} \langle 1 \vert
\bigg\},
\end{equation}
whose eigenvalues are given by
\begin{equation}
\label{eq:eigenvalues_ghz_fermion_2}
\lambda_{\text{GHZ}_1}^{F1} = \frac{e^{\pi \alpha \omega}}{2(1+e^{\pi \alpha \omega})}, \quad
\lambda_{\text{GHZ}_2}^{F1} = \frac{2+e^{\pi \alpha \omega}}{2(1+e^{\pi \alpha \omega})}.
\end{equation}
Substituting these eigenvalues into the AR $q$-conditional entropy, we obtain
\begin{equation}
\label{eq:ar_q_entropy_fermion}
S^{\text{GHZ},N}_{F,q} = \frac{1}{q-1} \Bigg\{
1 - \frac{1 + (1+2e^{\pi \alpha \omega})^q}{e^{
\pi \alpha \omega q} + (2+e^{\pi \alpha \omega})^q}
\Bigg\}.
\end{equation}

Similarly, the pure $N$-partite $W$ state [see Eq.~\eqref{WBF}] for the fermionic field, re-expressed in terms of the Unruh-diamond modes [Eqs.~\eqref{f-jt} and \eqref{f-jft}], is
\begin{equation}
\begin{split}
\lvert W^{F}_{123\cdots N+1}\rangle
&=
\frac{1}{\sqrt{N}}
\bigg\{
\lvert 1\rangle_{1}\lvert 0\rangle_{2}\cdots\lvert 0\rangle_{N-1}
\Bigl[
(1+e^{-\pi\alpha\omega})^{-\frac{1}{2}}
\lvert 0\rangle_{N,\mathrm{int}}\lvert 0\rangle_{N+1,\mathrm{ext}}\\
&+
(1+e^{\pi\alpha\omega})^{-\frac{1}{2}}
\lvert 1\rangle_{N,\mathrm{int}}\lvert 1\rangle_{N+1,\mathrm{ext}}
\Bigr]
+
\lvert 0\rangle_{1}\lvert 1\rangle_{2}\cdots\lvert 0\rangle_{N-1}
\Bigl[
(1+e^{-\pi\alpha\omega})^{-\frac{1}{2}}\\
&\times\lvert 0\rangle_{N,\mathrm{int}}\lvert 0\rangle_{N+1,\mathrm{ext}}
+
(1+e^{\pi\alpha\omega})^{-\frac{1}{2}}
\lvert 1\rangle_{N,\mathrm{int}}\lvert 1\rangle_{N+1,\mathrm{ext}}
\Bigr]
+\cdots\\
&+
\lvert 0\rangle_{1}\lvert 0\rangle_{2}\cdots\lvert 1\rangle_{N-1}
\Bigl[
(1+e^{-\pi\alpha\omega})^{-\frac{1}{2}}
\lvert 0\rangle_{N,\mathrm{int}}\lvert 0\rangle_{N+1,\mathrm{ext}}\\
&+
(1+e^{\pi\alpha\omega})^{-\frac{1}{2}}
\lvert 1\rangle_{N,\mathrm{int}}\lvert 1\rangle_{N+1,\mathrm{ext}}
\Bigr]
+
\lvert 0\rangle_{1}\lvert 0\rangle_{2}\cdots\lvert 0\rangle_{N-1}
\lvert 1\rangle_{N,\mathrm{int}}\lvert 0\rangle_{N+1,\mathrm{ext}}
\bigg\}.
\end{split}
\end{equation}
Tracing out the inaccessible exterior diamond modes, the reduced density operator $\rho^{F,W}_{123\cdots N_\mathrm{int}}$ reads
\begin{equation}
\rho^{F,W}_{123\cdots N_\mathrm{int}}
=
\frac{1}{N}
\bigl(
\rho^{F}_{\mathrm{diag}}
+
\rho^{F}_{\mathrm{off}}
\bigr).
\end{equation}
The diagonal part $\rho^{F}_{\mathrm{diag}}$ describes the classical population terms associated with single-excitation configurations, reading
\begin{align}
\rho^{F}_{\mathrm{diag}}
&=
\sum_{i=1}^{N-1}
\lvert 1\rangle_{i}\langle 1\rvert
\bigotimes_{j=1(j\neq i)}^{N-1}
\lvert 0\rangle_{j}\langle 0\rvert
\Bigl[
(1+e^{-\pi\alpha\omega})^{-1}
\lvert 0\rangle_{N,\mathrm{int}}\langle 0\rvert
\nonumber\\
&+
(1+e^{\pi\alpha\omega})^{-1}
\lvert 1\rangle_{N,\mathrm{int}}\langle 1\rvert
\Bigr]
+
\bigotimes_{i=1}^{N-1}
\lvert 0\rangle_{i}\langle 0\rvert
\lvert 1\rangle_{N,\mathrm{int}}\langle 1\rvert ,
\end{align}
while the off-diagonal part $\rho_F^{\mathrm{off}}$ arises from quantum coherence between different single-excitation sectors of the fermionic $W$ state and can be expressed as
\begin{equation}
\begin{split}
\rho^{F}_{\mathrm{off}}
&=
\sum_{\substack{i,j=1 \\ i\neq j}}^{N-1}
\lvert 1\rangle_{i}\langle 0\rvert
\lvert 0\rangle_{j}\langle 1\rvert
\bigotimes_{\substack{k=1\\k\neq i,j}}^{N-1}
\lvert 0\rangle_{k}\langle 0\rvert
\Bigl[
(1+e^{-\pi\alpha\omega})^{-1}\\
&\times\lvert 0\rangle_{N,\mathrm{int}}\langle 0\rvert
+
(1+e^{\pi\alpha\omega})^{-1}
\lvert 1\rangle_{N,\mathrm{int}}\langle 1\rvert
\Bigr]\\
&+
\sum_{i=1}^{N-1}
\lvert 1\rangle_{i}\langle 0\rvert
\bigotimes_{\substack{j=1 \\ j\neq i}}^{N-1}
\lvert 0\rangle_{j}\langle 0\rvert
(1+e^{-\pi\alpha\omega})^{-\frac{1}{2}}
\lvert 0\rangle_{N,\mathrm{int}}\langle 1\rvert\\
&+
\sum_{i=1}^{N-1}
\lvert 0\rangle_{i}\langle 1\rvert
\bigotimes_{\substack{j=1 \\ j\neq i}}^{N-1}
\lvert 0\rangle_{j}\langle 0\rvert
(1+e^{-\pi\alpha\omega})^{-\frac{1}{2}}
\lvert 1\rangle_{N,\mathrm{int}}\langle 0\rvert .
\end{split}
\end{equation}
The nonvanishing eigenvalues of the reduced density matrix $\rho^{F,W}_{123\cdots N_\mathrm{int}}$ are found to be
\begin{equation}
\lambda^{F}_{W_1}
=
\frac{N-1}{N(1+e^{\pi\alpha\omega})},
\qquad
\lambda^{F}_{W_2}
=
\frac{1+N e^{\pi\alpha\omega}}{N(1+e^{\pi\alpha\omega})}.
\end{equation}
Tracing out the first $N-1$ interior subsystems, we obtain the reduced density matrix corresponding to the 
$N$-th interior mode as
\begin{equation}
\rho^{F,W}_{N_\mathrm{int}}
=
\frac{1}{N}
\Bigl\{
(N-1)(1+e^{-\pi\alpha\omega})^{-1}
\lvert 0\rangle_{N,\mathrm{int}}\langle 0\rvert
+
\bigl[(N-1)(1+e^{\pi\alpha\omega})^{-1}+1\bigr]
\lvert 1\rangle_{N,\mathrm{int}}\langle 1\rvert
\Bigr\},
\end{equation}
whose eigenvalues are given by
\begin{equation}
\lambda^{F1}_{W_1}
=
\frac{(N-1)e^{\pi\alpha\omega}}{N(1+e^{\pi\alpha\omega})},
\qquad
\lambda^{F1}_{W_2}
=
\frac{N+e^{\pi\alpha\omega}}{N(1+e^{\pi\alpha\omega})}.
\end{equation}
Substituting these eigenvalues into the AR $q$-conditional entropy, we finally obtain
\begin{equation}\label{ar-q-w-f}
S^{W,N}_{F,q}
=
\frac{1}{q-1}
\left\{
1-
\frac{(N-1)^q+\bigl(1+N e^{\pi\alpha\omega}\bigr)^q}
{\bigl[(N-1)e^{\pi\alpha\omega}\bigr]^q+\bigl(N+e^{\pi\alpha\omega}\bigr)^q}
\right\}.
\end{equation}

\begin{figure}
\begin{minipage}[t]{0.5\linewidth}
\centering
\includegraphics[width=3.0in,height=5.2cm]{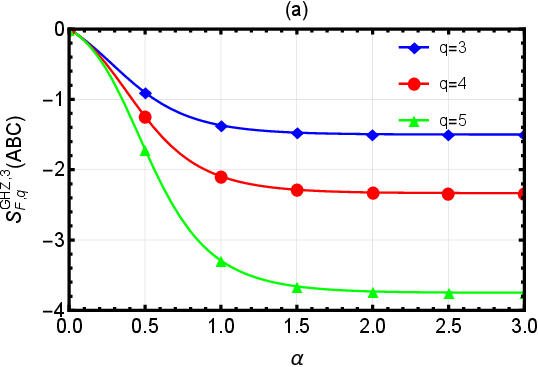}
\end{minipage}%
\begin{minipage}[t]{0.5\linewidth}
\centering
\includegraphics[width=3.0in,height=5.2cm]{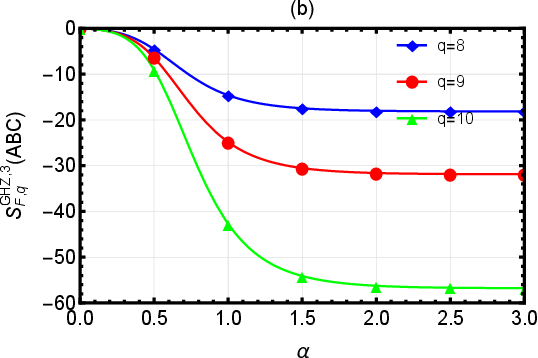}
\end{minipage}%
\vspace{0.5cm}
\begin{minipage}[t]{0.5\linewidth}
\centering
\includegraphics[width=3.0in,height=5.2cm]{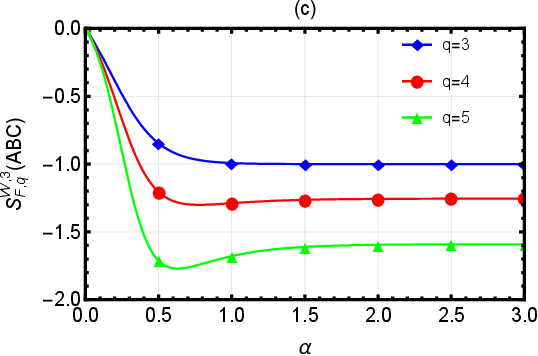}
\end{minipage}%
\begin{minipage}[t]{0.5\linewidth}
\centering
\includegraphics[width=3.0in,height=5.2cm]{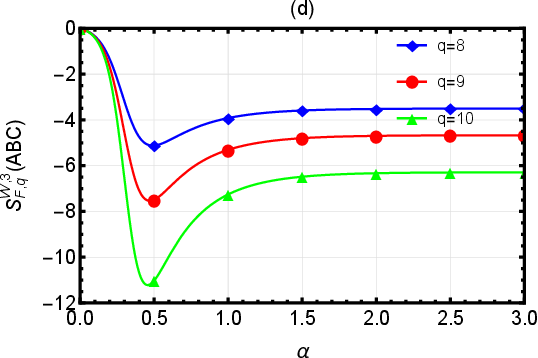}
\end{minipage}%
\caption{Dependence of the AR $q$-conditional entropy for fermionic $GHZ$ and $W$ states on the finite-lifetime parameter $\alpha$ for several values of $q$, with $\omega=1$ and $N=3$.}\label{fig:F-GHZ-W-3}
\end{figure}

In Fig.~\ref{fig:F-GHZ-W-3}, we plot the AR $q$-conditional entropies $S^{\mathrm{GHZ},3}_{F,q}(ABC)$ and $S^{\mathrm{W},3}_{F,q}(ABC)$ as a function of the finite-lifetime parameter $\alpha$ for different values of $q$, obtained from Eqs.~\eqref{eq:ar_q_entropy_fermion} and \eqref{ar-q-w-f}. The fermionic $GHZ$ state exhibits the same qualitative behavior as its bosonic counterpart: the AR $q$-conditional entropy decreases monotonically with increasing $\alpha$, indicating that the nonseparability is progressively enhanced as the causal-diamond temperature decreases. A qualitatively different behavior is observed for the fermionic $W$ state. Unlike the monotonic evolution of the bosonic $W$ state, $S^{\mathrm{W},3}_{F,q}(ABC)$ displays a pronounced nonmonotonic dependence on $\alpha$. As the observer's lifetime increases, the conditional entropy first decreases rapidly, reaches a minimum, and then approaches a finite asymptotic value. Consequently, the fermionic $W$ state exhibits a net enhancement of nonseparability relative to its initial value, demonstrating that causal-diamond-induced thermalization can reinforce, rather than merely degrade the nonseparability. Another notable feature is that the AR $q$-conditional entropies of both fermionic $GHZ$ and $W$ states remain negative for all values of $q$, including the limit $\alpha\rightarrow0$. This indicates that their nonseparability survives even for observers with arbitrarily short lifetimes, in sharp contrast to the bosonic case where the states become separable in the same limit. The enhanced robustness of fermionic states originates from Fermi-Dirac statistics, which suppress thermal excitations through the Pauli exclusion principle and thereby mitigate the degradation of the nonseparability under causal restrictions.

\begin{figure}
\begin{minipage}[t]{0.5\linewidth}
\centering
\includegraphics[width=3.0in,height=5.2cm]{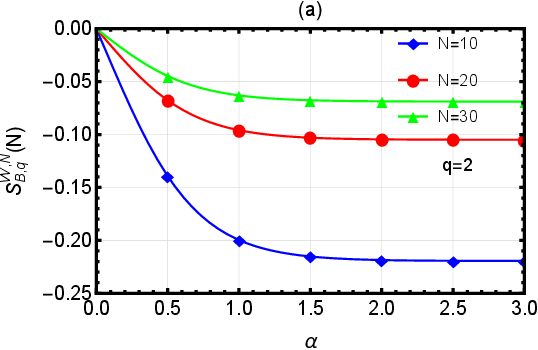}
\end{minipage}%
\begin{minipage}[t]{0.5\linewidth}
\centering
\includegraphics[width=3.0in,height=5.2cm]{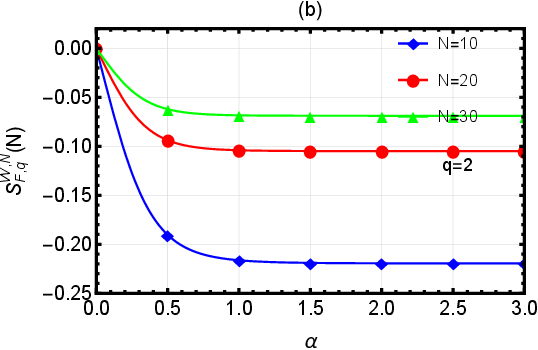}
\end{minipage}%
\vspace{0.3cm}
\begin{minipage}[t]{0.5\linewidth}
\centering
\includegraphics[width=3.0in,height=5.2cm]{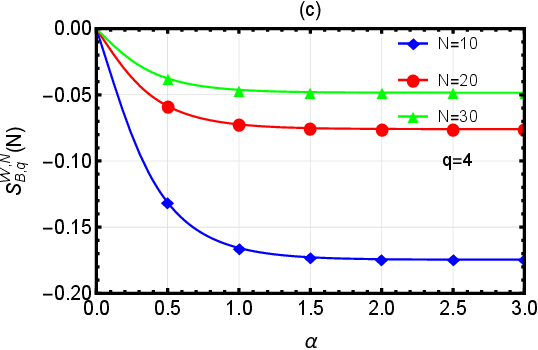}
\end{minipage}%
\begin{minipage}[t]{0.5\linewidth}
\centering
\includegraphics[width=3.0in,height=5.2cm]{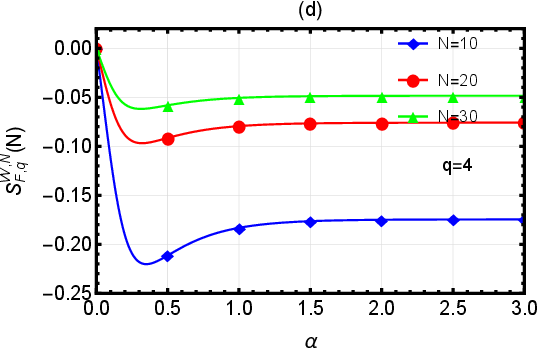}
\end{minipage}%
\vspace{0.3cm}
\begin{minipage}[t]{0.5\linewidth}
\centering
\includegraphics[width=3.0in,height=5.2cm]{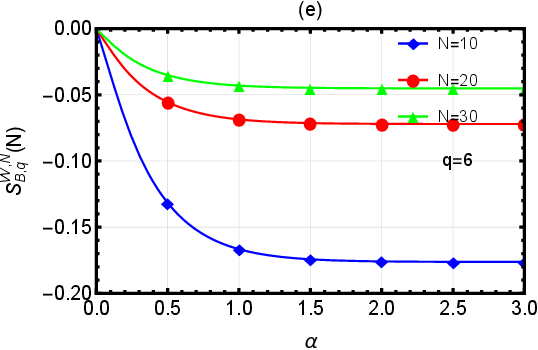}
\end{minipage}%
\begin{minipage}[t]{0.5\linewidth}
\centering
\includegraphics[width=3.0in,height=5.2cm]{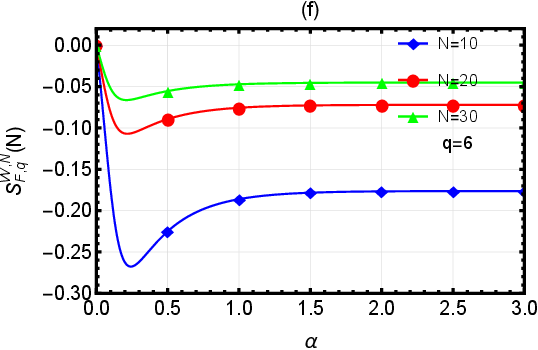}
\end{minipage}%
\vspace{0.3cm}
\begin{minipage}[t]{0.5\linewidth}
\centering
\includegraphics[width=3.0in,height=5.2cm]{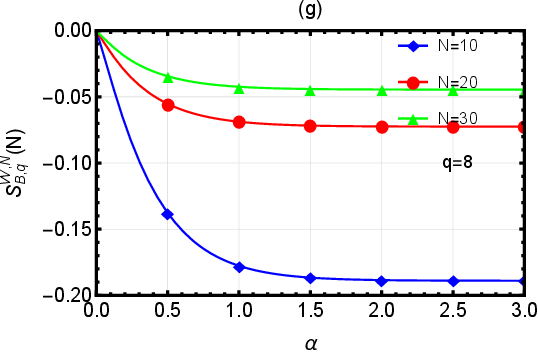}
\end{minipage}%
\begin{minipage}[t]{0.5\linewidth}
\centering
\includegraphics[width=3.0in,height=5.2cm]{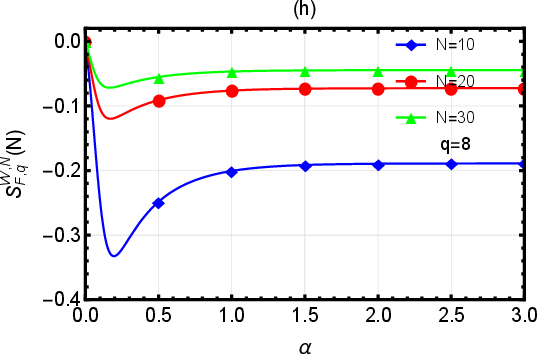}
\end{minipage}%
\caption{AR $q$-conditional entropy of both bosonic and fermionic fields for $W$ states in causal diamond spacetime, plotted as a function of the observer’s finite-lifetime parameter $\alpha$ for different values of $q$ and $N$, with $\omega = 1$.}\label{fig:BF-W-GHZ-N}
\end{figure}

The analytical expressions in Eqs.~\eqref{BGHZ1} and \eqref{eq:ar_q_entropy_fermion} show that the AR $q$-conditional entropy of the $GHZ$ state is independent of the total number of subsystems $N$ for both bosonic and fermionic fields. In contrast, the corresponding expressions for the $W$ state [Eqs.~\eqref{SBW} and \eqref{ar-q-w-f}] explicitly depend on $N$. To illustrate this feature, Fig.~\ref{fig:BF-W-GHZ-N} presents the AR $q$-conditional entropies $S^{W,N}_{B,q}$ and $S^{W,N}_{F,q}$ as a function of the finite-lifetime parameter $\alpha$ for different values of $N$ and $q$.
As shown in Fig.~\ref{fig:BF-W-GHZ-N}, the AR $q$-conditional entropy increases monotonically with increasing $N$ for both bosonic and fermionic $W$ states at fixed $\alpha$, indicating a gradual reduction of nonseparability as the system size increases. This result demonstrates that, unlike the $GHZ$ state, the nonseparability of the $W$ state is intrinsically sensitive to the number of subsystems in causal diamonds. The different scaling behaviors reflect the distinct entanglement structures of the two classes of multipartite states and show that causal-diamond thermalization affects the robustness of $W$ states more strongly as the system size increases.

\section{Conclusions}
We have studied the effects of causal-diamond-induced thermalization on the  nonseparability of $N$-partite $GHZ$ and $W$ states for both bosonic and fermionic fields using the AR $q$-conditional entropy. Initially, $N$ observers are assumed to share multipartite GHZ and $W$ states prepared in the global modes. We then consider a configuration in which $N-1$ subsystems remain inertial in the asymptotically flat region, while a single subsystem is associated with a finite-lifetime observer whose accessible region is restricted to the interior of a causal diamond. The presence of the diamond horizon induces a nontrivial decomposition of field modes into accessible and inaccessible sectors. Tracing over the causally disconnected region yields an effectively mixed state, from which we obtain analytical expressions for the $N$-partite AR $q$-conditional entropy characterizing  nonseparability under causal constraints. Our results reveal a clear distinction between bosonic and fermionic fields in causal diamond. Fermionic nonseparability is generally more robust against causal-diamond-induced thermalization than bosonic nonseparability. In the bosonic case, both GHZ and $W$ states exhibit a transition from nonseparability to separability in the limit of vanishing observer lifetime. We further find that $GHZ$ states exhibit stronger nonseparability than $W$ states under identical causal restrictions. The nonseparability of GHZ states is independent of the particle number $N$, whereas $W$ states display a pronounced $N$-dependence, with nonseparability decreasing as $N$ increases in diamond
spacetime. Most notably, we find that causal-diamond-induced thermalization can enhance the net nonseparability of fermionic $W$ states, in sharp contrast to the usual expectation of monotonic degradation of bosonic nonseparability under relativistic thermal effects. These results suggest that the robustness of nonseparability in relativistic settings is jointly determined by particle statistics, entanglement structure, and observer lifetime, highlighting the necessity of tailoring quantum resources to the underlying causal structure for relativistic quantum information tasks.

\begin{acknowledgments}
This work is supported by the National Natural Science Foundation of China (12575056) and LiaoNing Revitalization Talents Program (XLYC2503099).
\end{acknowledgments}

$\textbf{Data Availability Statement}$

This manuscript has no associated data.

$\textbf{Conflict of interest}$

The authors declare no conflicts of interest.

\appendix


\begin{thebibliography}{99}
\bibitem{SDF1}
W. Liu, C. Wen, J. Wang, Lorentz violation alleviates gravitationally induced entanglement degradation, J. High Energy Phys. {\bf2025}, 184 (2025).

\bibitem{SDF2}
S. Sen, A. Mukherjee and S. Gangopadhyay, Entanglement degradation as a tool to detect signatures of modified gravity, Phys. Rev. D {\bf109}, 046012 (2024).

\bibitem{SDF3}
I. Fuentes-Schuller and R. B. Mann, Alice falls into a black hole: Entanglement in non-inertial frames, Phys. Rev. Lett. {\bf95}, 120404 (2005).

\bibitem{SDF4}
Q. Pan and J. Jing, Hawking radiation, entanglement, and teleportation in the background of an asymptotically flat static black hole, Phys. Rev. D {\bf78}, 065015 (2008).

\bibitem{SDF5}
H. Wu and L. Chen, Orbital angular momentum entanglement in noninertial reference frame, Phys. Rev. D {\bf107}, 065006 (2023).

\bibitem{SDF6}
S. M. Wu, X. W. Fan, R. D. Wang, H. Y. Wu, X. L. Huang and H. S. Zeng, Does Hawking effect always degrade fidelity of quantum teleportation in Schwarzschild spacetime?, J. High Energy Phys. {\bf2023}, 232 (2023).


\bibitem{SDF7}
E. Mart\'{\i}n-Mart\'{\i}nez, L. J. Garay and J. Le\'{o}n, Unveiling quantum entanglement degradation near a Schwarzschild black hole,  Phys. Rev. D {\bf82}, 064006 (2010).

\bibitem{SDF8}
P. M. Alsing, I. Fuentes-Schuller, R. B. Mann and T. E. Tessier, Entanglement of Dirac fields in non-inertial frames, Phys. Rev. A {\bf74}, 032326 (2006).

\bibitem{SDF9}
W. M. Li, S. M. Wu, Bosonic and fermionic coherence of N-partite states in the background of a dilaton black hole,  J. High Energy Phys. {\bf2024}, 144 (2024).

\bibitem{SDF10}
M. M. Du, H. W. Li, S. T. Shen, X. J. Yan, X. Y. Li, L. Zhou, W. Zhong and Y. B. Sheng, Maximal steered coherence in the background of Schwarzschild space-time, Eur. Phys. J. C {\bf84}, 450 (2024).

\bibitem{SDF11}
S. Elghaayda, A. Ali, S. Al-Kuwari and M. Mansour, Physically accessible and inaccessible quantum correlations of Dirac fields in Schwarzschild spacetime, Phys. Lett. A {\bf525}, 129915 (2024).


\bibitem{SDF12}
Y. Xiong, Z. Pi, T. Zhang, X. Huang, Quantum discord of mixed states under noisy channels in the
curved spacetime, Eur. Phys. J. C  {\bf86}, 317 (2026).


\bibitem{SDF60}
X. Liu, C. Zeng, J. Wang, Generation of quantum entanglement in superposed diamond spacetime, Eur. Phys. J. C {\bf85},  539  (2025).


\bibitem{SDF61}
H. E. Camblong, A. Chakraborty, P. Lopez-Duque, C. R. Ord\'{o}\~{n}ez, Entanglement degradation in causal diamonds,
Phys. Rev. D {\bf109}, 105003 (2024).



\bibitem{SDF13}
A. Ali, S. Al-Kuwari, M. Ghominejad, M. T. Rahim, D. Wang and S. Haddadi, Quantum characteristics near event horizons, Phys. Rev. D {\bf110}, 064001 (2024).

\bibitem{SDF14}
C. Y. Liu, Z. W. Long and Q. L. He, Quantum coherence and quantum Fisher information of Dirac particles in curved spacetime under decoherence, Phys. Lett. B {\bf857}, 138991 (2024).

\bibitem{SDF15}
T. Zhang, X. Wang and S. M. Fei, Hawking effect can generate physically inaccessible genuine tripartite nonlocality, Eur. Phys. J. C {\bf83}, 607 (2023).

\bibitem{SDF16}
S. M. Wu, H. Y. Wu, Y. X. Wang, J. Wang, Gaussian tripartite steering in Schwarzschild black hole, Phys. Lett. B {\bf865},  139493 (2025).

\bibitem{SDF17}
G. Adesso, I. Fuentes-Schuller and M. Ericsson, Continuous variable entanglement sharing in non-inertial frames, Phys. Rev. A {\bf76}, 062112 (2007).

\bibitem{SDF18}
P. K. Dahal, K. Hymas, Spatiotemporal entanglement of the vacuum, Phys. Rev. A {\bf113}, L020201 (2026).



\bibitem{SDF19}
G. W. Mi, X. Huang, S. M. Fei, T. Zhang, Quantumness near the Schwarzschild black hole based on W-state,
Ann. Phys. (Berlin)  {\bf10}, 1002 (2025).

\bibitem{SDF20}
S. Harikrishnan, S. Jambulingam, P. P. Rohde and C. Radhakrishnan, Accessible and inaccessible quantum coherence in relativistic quantum systems, Phys. Rev. A {\bf105}, 052403 (2022).

\bibitem{SDF21}
G. W. Mi, X. Huang, S. M. Fei, T. Zhang, Genuine four-partite Bell nonlocality in the curved spacetime, Eur. Phys. J. C {\bf85}, 354, (2025).


\bibitem{SDF22}
T. Y. Wang and D. Wang, Entropic uncertainty relations in Schwarzschild space-time, Phys. Lett. B {\bf855}, 138876 (2024).

\bibitem{SDF23}
P. Michalski, A, Dragan, Detection of quantum entanglement across the event horizon, Phys. Rev. A {\bf 114}, 012420 (2026).

\bibitem{SDF24}
G. W. Mi, X. Huang, S. M. Fei, T. Zhang, Impact of the Hawking Effect on the fully entangled
fraction of three-qubit states in Schwarzschild spacetime, Ann. Phys. (Berlin) {\bf537}, 2400308
(2024).

\bibitem{SDF25}
R. J. Yao, D. Wang, Generalized entropic uncertainty relation and nonclassicality for Schwarzschild black holes, Phys. Rev. D {\bf113}, 065002 (2026).

\bibitem{SDF26}
S. M. Wu, X. W. Teng, J. X. Li, S. H. Li, T. H. Liu, J. C. Wang, Genuinely accessible and inaccessible
entanglement in Schwarzschild black hole. Phys. Lett. B {\bf848}, 138334 (2024).

\bibitem{SDF27}
A. Belfiglio, O. Luongo, S. Mancini, Quantum entanglement in cosmology, Phys. Rep. {\bf1146}, 1 (2025).


\bibitem{SDF28}
X. Liu, W. Liu, Z. Liu, J. Wang, Harvesting correlations from BTZ black hole coupled to a Lorentz-violating vector field, J. High Energy Phys. {\bf2025}, 94 (2025).


\bibitem{SDF29}
Z. D. Wei, W. Han, Y. J. Zhang, Z. X. Man, Y. J. Xia, H. Fan, Effect of the gravitational redshift on the precision of phase estimation, Phys. Rev. D \textbf{111},  026007 (2025).


\bibitem{SDF30}
W. Izquierdo,  J. Beltran,  E. Arias, Enhancement of harvesting vacuum entanglement in Cosmic String Spacetime, J. High Energy Phys. {\bf2025}, 049 (2025).


\bibitem{SDF31}
Z. Liu, W. Liu, X. Liu, J. Wang, Wormhole-Induced correlation: A Link Between Two Universes, Phys. Lett. B {\bf879}, 140667  (2026).

\bibitem{SDF32}
S. M. Wu, C. X. Wang, D. D. Liu, X. L. Huang, H. S. Zeng, Would quantum coherence be increased by curvature effect in de Sitter space?, J. High Energy Phys. {\bf2023}, 115  (2023).

\bibitem{SDF33}
A. Chakraborty,  L. Hackl,  M. Zych, Entanglement harvesting in quantum superposed spacetime, Phys. Rev. D {\bf111}, 104052 (2025).

\bibitem{SDF34}
Z. Liu, R. Q. Yang, H. Fan, J. Wang, Simulation of the massless Dirac field in 1+1D curved spacetime, Sci. China Phys. Mech. Astron. {\bf68}, 290411 (2025).

\bibitem{SDF35}
S. M. Wu, R. D. Wang, X. L. Huang, Z. Wang, Does gravitational wave assist vacuum steering and Bell nonlocality?, J. High Energy Phys. {\bf2024}, 155 (2024).

\bibitem{SDF36}
Z. Liu, Y. Li, Z. Tian, J. Wang, Scrambling-Enhanced Quantum Battery Charging in Black Hole Analogues, Adv. Sci. {\bf 11}, e20281 (2026).


\bibitem{SDF37}
X. Liu, Z. Tian, J. Jing, Entanglement dynamics in $\kappa$-deformed
spacetime, Sci. China Phys. Mech. Astron. {\bf67}, 100411 (2024).


\bibitem{SDF38}
S. Barman, I. Chakraborty, S. Mukherjee, Signatures of gravitational wave memory in the radiative process of entangled quantum probes, Phys. Rev. D {\bf111}, 025021 (2025).

\bibitem{SDF39}
A. J. Torres-Arenas, Q. Dong, G. H. Sun, W. C. Qiang and S. H. Dong, Entanglement measures of W-state in noninertial frames, Phys. Lett. B {\bf789}, 93 (2019).


\bibitem{SDF40}
C. Y. Liu, Z. W. Long, Q. L. He,  Would the fidelity of quantum teleportation be increased by a local filtering operation near a dilaton black hole under decoherence?, Eur. Phys. J. C {\bf85}, 926 (2025).


\bibitem{SDF41}
F. Shahbazi, S. Haseli, H. Dolatkha, and S. Salimi, Entropic uncertainty relation in Garfinkle-Horowitz-Strominger dilation black hole, JCAP {\bf10}, 047 (2020).


\bibitem{SDF42}
J. Foo, C. S. Arabaci, M. Zych, R. B. Mann, Quantum Signatures of Black Hole Mass Superpositions, Phys. Rev. Lett. {\bf129}, 181301  (2022).


\bibitem{SDF43}
A. A. A. Filho, W. Liu, Entanglement, equivalence principle, and HBAR entropy, in a new bumblebee
black hole, arXiv:2512.17567.


\bibitem{SDF44}
M. M. Du, H. W. Li, Z. Tao, S. T. Shen, X. J. Yan, X. Y. Li, W. Zhong, Y. B. Sheng and L. Zhou, Basis-independent quantum coherence and its distribution under relativistic motion, Eur. Phys. J. C {\bf84}, 838 (2024).

\bibitem{SDF45}
Y. Ji, J. Zhang and H. Yu, Entanglement harvesting in cosmic string spacetime, J. High Energy Phys. {\bf2024}, 161 (2024).

\bibitem{SDF46}
A. A. Svidzinsky, M. O. Scully,  W. Unruh, Minkowski vacuum entanglement and accelerated oscillator chains, Phys. Rev. D {\bf111}, 045022 (2025).

\bibitem{SDF47}
Q. Liu, T. Liu, C. Wen, J. Wang, Optimal quantum strategy for locating Unruh channels, Phys. Rev. A {\bf110}, 022428 (2024).

\bibitem{SDF48}
Y. K. Zhang, L. J. Li, X. K. Song, L. Ye and D. Wang, Entropic uncertainty and quantum non-classicality of Unruh-Dewitt detectors in relativity, Phys. Lett. B {\bf858}, 139063 (2024).

\bibitem{SDF49}
S. H. Li, S. H. Shang, S. M. Wu, Does acceleration always degrade quantum entanglement for tetrapartite Unruh-DeWitt detectors?, J. High Energy Phys. {\bf2025}, 214  (2025).


\bibitem{SDF50}
T. Gonzalez-Raya, S. Pirandola and M. Sanz, Satellite-based entanglement distribution and quantum teleportation with continuous variables, Commun. Phys. {\bf7},  126 (2024).


\bibitem{SDF51}
Y. Tang, W. Liu, Z. Liu, J. Wang, Can the signatures of quantum superposition be detected
through correlation harvesting?, J. High Energy Phys. {\bf2026}, 045 (2026). 

\bibitem{SDF52}
Y. Tang, W. Liu, J. Wang, Observational signature of Lorentz violation in acceleration radiation, Eur.
Phys. J. C {\bf 85}, 1108 (2025).

\bibitem{SDF53}
G. Ciliberto, S. Emig, N. Pavloff, M. Isoard, Violation of Bell inequalities in an analog black hole, Phys. Rev. A {\bf109}, 063325 (2024).

\bibitem{SDF54}
J. K. Basak, D. Giataganas, S. Mondal and W. Y. Wen, Reflected entropy and Markov gap in noninertial frames, Phys. Rev. D {\bf108}, 125009 (2023).



\bibitem{SDF55}
B. Yu, X. Y. Yang, X. Hu, Z. X. Jin, X. Huang, Nonlocal advantage of quantum imaginarity in Schwarzschild spacetime, arXiv:2604.03633


\bibitem{SDF56}
S. M. Wu and H. S. Zeng, Genuine tripartite nonlocality and entanglement in curved spacetime, Eur. Phys. J. C {\bf82}, 4 (2022).

\bibitem{SDF57}
F. Ming, Z. Q. Xu, T. T. Lu, L. Dong, B. L. Fang, X. Hu, Y. Yu, H. Yang, D. Wang, Unveiling gravity-induced quantumness by three-measurement uncertainty relations. Eur. Phys. J. C {\bf86}, 85 (2026).


\bibitem{SDF58}
I. Agullo, A. Delhom, \'A. Parra-L\'opez, Toward the observation of entangled pairs in BEC analog expanding universes, Phys. Rev. D \textbf{110}, 125023 (2024).


\bibitem{SDF59}
Q. Xiao, Y. Chen, T. Liu, Effects of Lorentz symmetry breaking on quantum coherence in an expanding universe,
Phys. Lett. B {\bf872}, 140133  (2026).


\bibitem{QFL1}
S. Mondal, W. Y. Wen, Entanglement at the soft-hair horizon, Phys. Lett. B {\bf833},  137385 (2022).



\bibitem{QFL2}
Y. Xu, Y. Chen, Q. Xiao, X. Gong, Nonlocal correlations of fermionic entanglement in the spacetime of Einstein-Gauss-Bonnet black hole,  Eur. Phys. J. C {\bf86}, 915 (2026).


\bibitem{QFL3}
F. L. Lin, S. Mondal, Can Newtonian Gravity Produce Quantum Entanglement?, Phys. Rev. D {\bf113}, L061901 (2026).




\bibitem{SDF62}
D. Bluvstein, A. A. Geim, S. H. Li \textit{et al.}, A fault-tolerant neutral-atom architecture for universal quantum computation, Nature \textbf{649}, 39 (2026).


\bibitem{SDF63} D. Gao, D. Fan, C. Zha, J. Bei, G. Cai, J. Cai, S. Cao, F. Chen, J. Chen, \textit{et al.}, Establishing a New Benchmark in Quantum Computational Advantage with 105-qubit Zuchongzhi 3.0 Processor, Phys. Rev. Lett. \textbf{134}, 090601 (2025).


\bibitem{SDF64} A. I. Khinchin, Mathematical Foundations of Information Theory (Dover, New York, 1957).


\bibitem{SDF65} C. Tsallis, R. S. Mendes, and A. R. Plastino, The role of constraints within generalized nonextensive statistics, Physica A \textbf{261}, 534 (1998).


\bibitem{SDF66} J. F. Bercher, Some properties of generalized Fisher information in the context of nonextensive thermostatistics, Physica A \textbf{392}, 3140 (2013).


\bibitem{SDF67} C. Tsallis, Possible generalization of Boltzmann-Gibbs statistics, J. Stat. Phys. \textbf{52}, 479 (1988).


\bibitem{SDF68} P. Martinetti, C. Rovelli, Diamond's temperature: Unruh effect for bounded trajectories and thermal time hypothesis, Class. Quantum Gravity \textbf{20}, 4919 (2003).


\bibitem{SDF69} D. Ida, T. Okamoto, M. Saito, Modular theory for operator algebra in a bounded region of space-time and quantum entanglement, Prog. Theor. Exp. Phys. \textbf{2013}, 083E03 (2013).


\bibitem{SDF70} D. Su, T. Ralph, Spacetime diamonds, Phys. Rev. D \textbf{93}, 044023 (2016).


\bibitem{SDF71} A. Chakraborty, H. Camblong, C. Ordóñez, Thermal effect in a causal diamond: Open quantum systems approach, Phys. Rev. D \textbf{106}, 045027 (2022).


\bibitem{SDF72} P. Francesco, P. Mathieu, D. Sénéchal, \textit{Conformal Field Theory}, (Springer Science \& Business Media, Berlin, 2012).


\bibitem{SDF73} P. D. Hislop, R. Longo, Modular structure of the local algebras associated with the free massless scalar field theory, Commun. Math. Phys. \textbf{84}, 71 (1982).


\bibitem{SDF74} P. Di Francesco, P. Mathieu, and D. S\'{e}n\'{e}chal, \textit{Conformal Field Theory}, (Springer, New York, 1997).


\bibitem{SDF75} N. D. Birrell, P. C. W. Davies, \textit{Quantum Fields in Curved Space}, (Cambridge University Press, Cambridge, England, 1984).


\bibitem{SDF76} S. Takagi, Vacuum noise and stress induced by uniform acceleration: Hawking-Unruh effect in Rindler manifold of arbitrary dimension, Prog. Theor. Phys. Suppl. \textbf{88}, 1 (1986).


\bibitem{SDF77} L. C. Crispino, A. Higuchi, G. E. Matsas, The Unruh effect and its applications, Rev. Mod. Phys. \textbf{80}, 787 (2008).














\end{thebibliography}
\end{document}